\documentclass[%
aps,
prapplied,
superscriptaddress,
longbibliography,
 amsmath,amssymb,
 reprint,%
]{revtex4-2}

\usepackage{hyperref}

\usepackage[capitalise]{cleveref}

\usepackage{graphicx}
\usepackage{dcolumn}
\usepackage{bm}
\usepackage{float}
\usepackage{array}
\usepackage{textcomp}
\usepackage{physics}
\usepackage[dvipsnames]{xcolor}
\usepackage{multirow}
\usepackage{booktabs}
\usepackage{upgreek}
\usepackage{xr}
\usepackage{cleveref}
\usepackage{tabularx}
\usepackage{amsmath}
\usepackage{appendix}
\usepackage{gensymb}
\usepackage{fixltx2e}
\usepackage{xcolor}
\usepackage{newtxtext}
\usepackage{newtxmath}

\makeatletter
\DeclareRobustCommand{\change}{%
  \@bsphack
  \normalcolor
  \@esphack
}
\DeclareRobustCommand{\stopchange}{%
  \@bsphack
  \normalcolor
  \@esphack
}
\makeatother

\begin{document}
\newcolumntype{C}[1]{>{\centering\arraybackslash}p{#1}}
\newcommand{\beginsupplement}{%
        \setcounter{table}{0}
        \renewcommand{\thetable}{S\arabic{table}}%
        \setcounter{figure}{0}
        \renewcommand{\thefigure}{S\arabic{figure}}%
     }
     
\preprint{AIP/123-QED}

\title[]{Alternating-Bias Assisted Annealing of Amorphous Oxide Tunnel Junctions}
\author{David~P.~Pappas}
\author{Mark~Field}
\author{Cameron~J.~Kopas}
\author{Joel~A.~Howard}
\author{Xiqiao~Wang}
\author{Ella~Lachman}
\author{Jinsu~Oh}
\author{Lin~Zhou}
\author{Alysson~Gold}
\author{Gregory~M.~Stiehl}
\author{Kameshwar~Yadavalli}
\author{Eyob~A.~Sete}
\author{Andrew~Bestwick}
\author{Matthew~J.~Kramer} 
\author{Josh~Y.~Mutus}

\date{\today} 
\begin{abstract}
\change 
Superconducting  quantum bits rely on ultra-thin, amorphous oxide tunneling barriers that can have significant inhomogeneities and defects as grown. This can result in relatively large uncertainties and deleterious effects in the circuits, limiting the scalability. Finding a robust solution to the junction reproducibility problem has been a long-standing goal in the field. Here, we demonstrate a transformational technique for controllably tuning the electrical properties of aluminum-oxide tunnel junctions. This is accomplished using a low-voltage, alternating-bias applied individually to the tunnel junctions, with which resistance tuning by more than 70\% can be achieved. The data indicates an improvement of coherence and reduction of two-level system defects. Transmission electron microscopy shows that the treated junctions are predominantly amorphous, albeit with a more uniform distribution of alumina coordination across the barrier. This technique is expected to be useful for other devices based on ionic amorphous materials.\stopchange
\end{abstract}

\maketitle

\section*{Introduction}
The preparation of amorphous thin film materials is surprisingly simple. While devices incorporating these materials are ubiquitous, their morphology is surprisingly complex and challenging to control. Among other similar materials, amorphous aluminum oxide   (a-AlO$_x$) tunnel    junctions stand out due to their importance in both cryogenic and room temperature devices. These applications range from those that utilize the Josephson effect \cite{JOSEPHSON1962PRL}, such as SQUID magnetometers \cite{Clarke2004},  superconducting qubits \cite{Devoret1997, Nakamura1999}, amplifiers \cite{JPA2009, TWPA2015}, RSFQ logic \cite{RSFQ1991}, to room temperature magnetic tunnel junctions \cite{ShajilNair2023} and memory elements \cite{ZHU2006, BEEM_RippardPRL2002, AlOxBreakdownJAP2004}. 
\begin{figure}[t]
\includegraphics[width=\columnwidth]{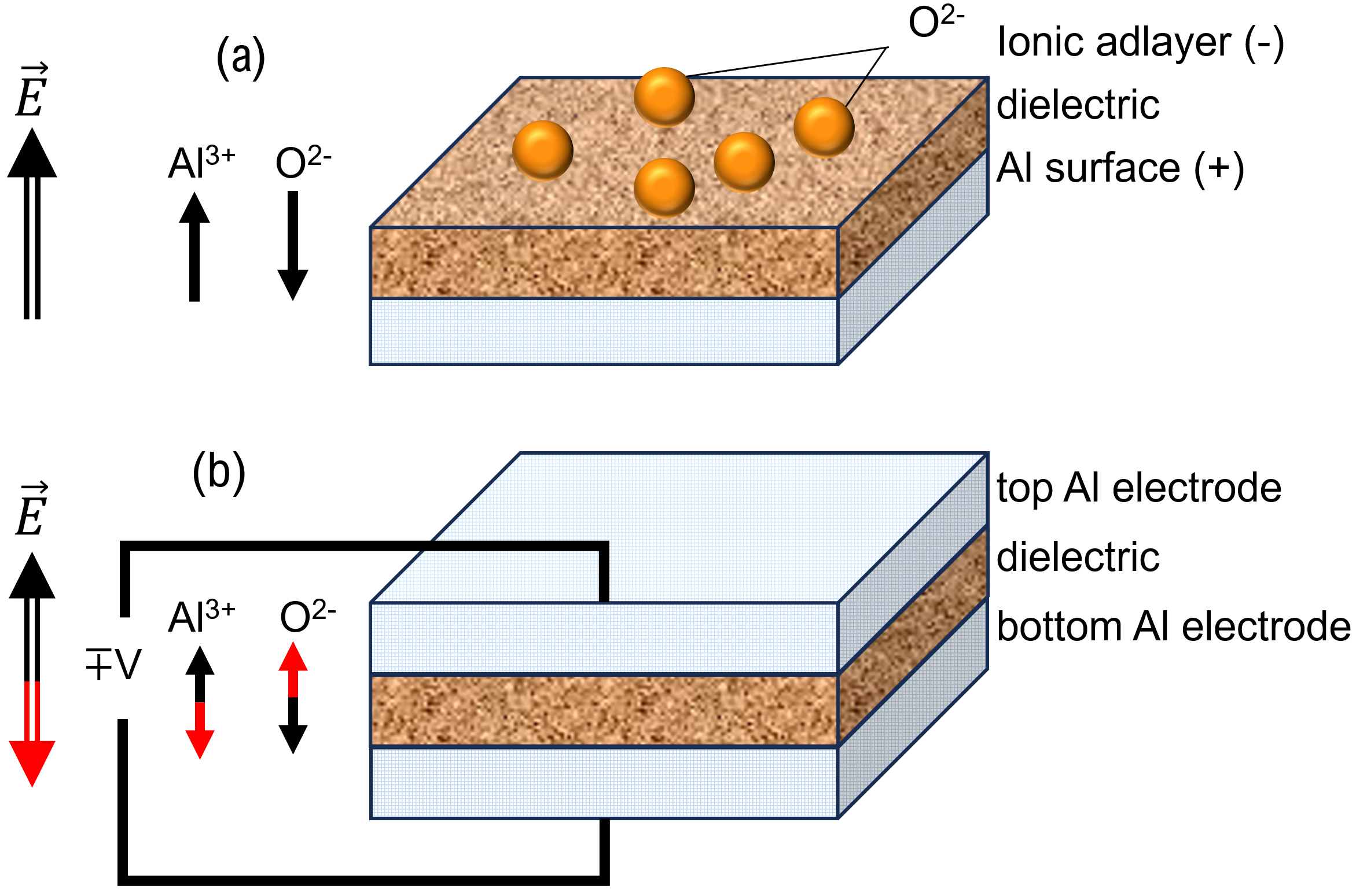}
\caption{\change\\Title: Models for thin-film oxidation of metals and ABAA\\
Legend: Panel(a) Static Cabrera-Mott Model:  Oxide growth on aluminum mechanism. The primary assumption is that adsorbed oxygen dissociates and ionizes. This creates a voltage $V_{Mott}$ relative to the Al, causing a strong electric field in the thin oxide. Due to the predominantly ionic-bonded nature of these films (56\%), aluminum cations and oxygen anions feel opposite forces due to this field. They exchange places, mediated by vacancy motion.\\ 
Panel (b): Dynamic Cabrera-Mott Effect (ABAA): The Al/AlO$_x$/Al trilayer is effectively mixed by applying alternating voltages onto the tunnel barrier electrodes. The voltage is on the order of magnitude of $V_{\mathrm{Mott}}$ but less than the dielectric breakdown voltage, between the two plates. This applies alternating and opposing forces on the two species.\stopchange}
\label{Fig:CabreraMottVsABAA}
\end{figure} \change 
It has traditionally been assumed that the structural and transport properties of these amorphous materials are frozen-in once they are grown, and that high temperature annealing is required in order to modify them. Here, we demonstrate an alternating-voltage based technique, i.e.”Alternating-Bias Assisted Annealing” (ABAA), to re-order the atoms in these materials at relatively low temperatures in order to tune the electrical properties. Superconducting devices are used \stopchange to demonstrate the efficacy of this technique. This is because, due to the cryogenic application of these oxide barriers for Josephson junctions, we can probe not only their resistance with high accuracy (using frequency), but also the prevalence of defects and loss using superconducting qubits. This is especially relevant to the field of quantum information since a-AlO$_x$ is a key ingredient in transmon qubits \cite{koch2007charge}, the invention of which has fueled much of the rapid development of superconducting applications research over the past decade. The junction properties are crucial in the overall properties and performance of the qubit. Most importantly, the critical current, $I_c$ of the barrier sets the Josephson energy of the device,  $E_J=I_c\Phi_0/2\pi$, where $\Phi_0$ is the flux quantum . $E_J$ determines the parameters of the transmon    such as the   frequency and charge noise sensitivity.      $I_c$ can be predicted by the Ambegaokar–Baratoff formula, i.e.  $I_C=\pi\Delta/2eR_n$~\cite{VanDuzer}, where, the variables $R_n$ and $\Delta$ are the resistance of the junction (in its normal state) and the superconducting gap of the electrodes. 

Unfortunately, while it is essential that $I_c$, and hence $R_n$,  is well defined, a large spread of $R_n$ in nominally identical junctions is typically observed, on the order of 1-10\% at the wafer level \cite{Muthusubramanian2023}. This is due to both intrinsic properties and extrinsic effects. Intrinsic issues include the exponential dependence of the tunneling on the thickness \cite{Zeng2016}; grain structure; and interface roughness. Extrinsic effects arise from the effects of lithography uncertainties \cite{InsituBandaid_Bilmes_2021}, processing, electrostatic sensitivity, and circuit design. 

\begin{figure*}
\includegraphics[width=\textwidth]{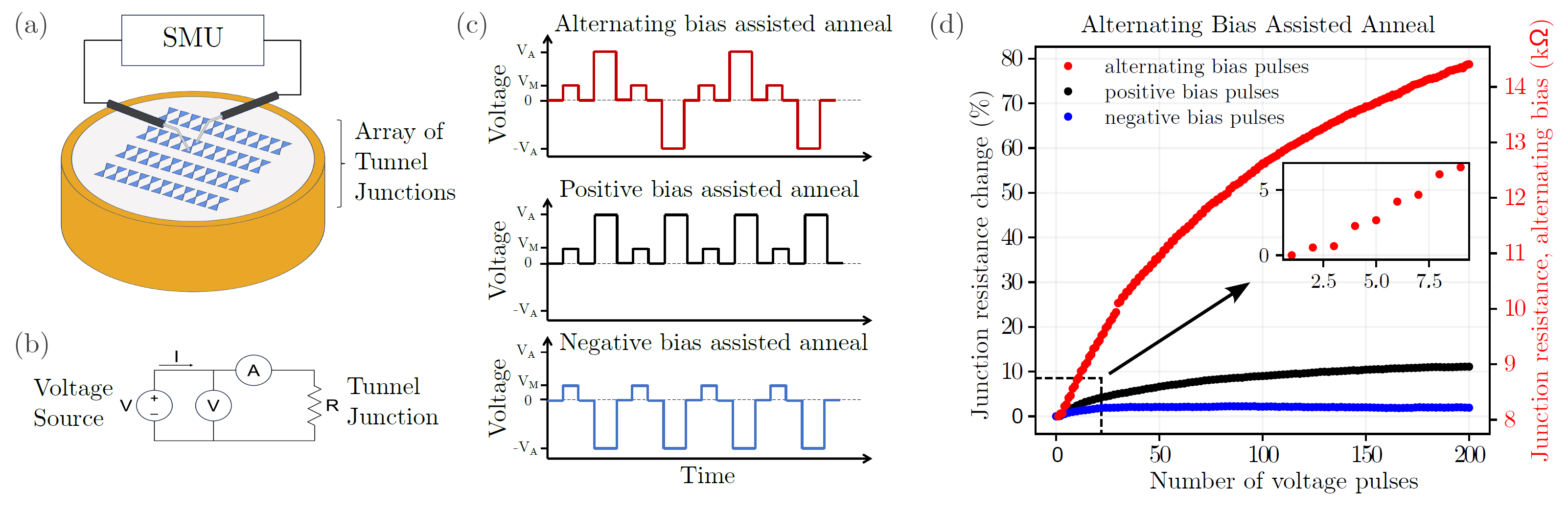}
\caption{\change\\Title: Schematic diagram of the bias schemes used to anneal tunnel junctions. \\
Legend: Panel (a): A block diagram of the apparatus. In all cases, the voltage source has the positive terminal connected to the top electrode and the negative terminal connected to the bottom electrode sitting on the substrate. The chuck can optionally be temperature controlled to assist with tuning. Typically 2-point probes are used to land on the individual qubit capacitor pads, since the junction resistance in this case is much higher than the probes. The junctions are formed on intrinsic-Si with ground rings around each qubit to avoid cross-talk.\\ Panel (b): Circuit diagram for the ABAA technique. \\
Panel (c): Bias pulses at $\mathrm{V_{A}}$ are applied for 1 second. In this paper $\mathrm{V_{A}}$ = 0.9 V. Between the bias pulses a low measurement pulse of $\mathrm{V_{M}}$ = 10 mV is used to measure the low bias resistance of the junction.\\
Panel (d): Data from three different junctions to illustrate voltage-assisted annealing using alternating bias assisted anneal (ABAA)  at $80 \degree \mathrm{C}$ vs. unipolar (positive and negative w.r.t. the top electrode) voltage pulses. The maximum voltage amplitude $V_{A} = \pm 0.9\mathrm{V}$ with a 1-second pulse length. The area of the junction under test is $1.55~ \mu\mathrm{m} \times 0.24~\mu \mathrm{m} = 0.37 ~(\mu\mathrm{m})^2$. The inset is a zoom-in on the start of the ABAA process and shows the cyclic change in junction resistance with a repeated sequence of small and large resistance changes when applying alternating bias bipolar pulses. The uncertainty in the resistance measurements is less than the size of the points.\stopchange}
\label{Fig:Bias_Schemes}
\end{figure*}
Moreover, the amorphous nature of the oxide layer results in the inclusion of defects that can cause two-level systems (TLS's) near the qubit frequency that typically reduce coherence, or the defects can result in two-level fluctuators (TLF's) at very low frequencies that push the qubit frequency around \cite{martinis2005decoherence,SchlorDecoherencePRL}.

The default method of making junctions, also used in this work, employs thermal oxidation of an aluminum   electrode surface and shadow evaporation. \cite{Lecocq2011} The accepted model for the oxidation process is described by the Cabrera-Mott model, \change illustrated in Figure 1(a), \stopchange in which the aluminum/oxygen kinetics in these ionically bonded materials \cite{Pauling1931}    is driven by an intrinsic electric field~\cite{OxidationReview, Cabrera_1949}. This process is convenient and powerful because it is conformal and self-limiting in thickness. In the thin-barrier limit the formation of the oxide barrier is governed by drift-dominated atomic diffusion \cite{OxidationReview, OxidationKinetics2012} of the charged constituent elements, Al$^{3+}$ and O$^{2-}$, in opposite directions, as illustrated in Fig.~\ref{Fig:CabreraMottVsABAA}~(a). This diffusion is driven by the strong $E$-field $\sim$~1~GV/m generated by the adsorbed ionic ad-layer.  In these, and other metal-insulator-metal oxides, the relevant parameter is referred to as the Mott voltage, $V$\textsubscript{Mott}$\approx$~0.5-1~V for aluminum oxide \cite{TemperatureAndPressureMottPotentials2012}.  

There have been many efforts to improve the properties of the a-AlO$_x$ material, focused primarily on understanding and optimizing the junction formation and post-processing, i.e. pressure, time, \cite{Mallison1995}, interface preparation, etches, simulations, etc., see Ref.~\cite{AlOxJunctionsLordiAPL2020, AlOxSimulationsCysterNPJQI2021, ImprovingJunctionReproducibilityPishchimovaSciReports2023} and citations therein. These efforts have met with limited success, and it is typically assumed that the properties of the junctions, as formed, are relatively frozen-in and difficult to adjust. For example,  annealing the   entire wafer to high temperature after defining the junctions can significantly change the resistances, however, this approach has limited efficacy\cite{AnnealingWafersNbAlAlOxNbVettoliere2020}. This is due to a highly variable response of the various junctions, presumably due to nanoscopic morphology differences of nominally identical junctions at the domain and atomic level. Subsequently, methods of annealing individual junctions have been developed~\cite{LocalizedAnnealingNbAlAlOxNbGranata2007,Hertzberg2021, LasiQScience2022, LasiQBerkeley2022,ebeamAnnealing, LaserAnnealDelft_1}. These methods typically are implemented by targeting a slightly lower resistance in fabrication and then locally heating the individual junctions with high-power lasers or an electron beam to bring them into tolerance.  However, these procedures involve the addition of complicated optical and imaging equipment     to the test and measurement apparatus,and appear to be limited to adjustments of the resistances up to about 10-12\% in Al/a-AlO$_x$/Al junctions, i.e. on the order of the junction spread. This limits the number of junction arrays that can be successfully adjusted, based on the initial distribution of the junctions.

On the other hand, it has been shown that oxidation can be enhanced by an external voltage   during growth (by e-beam bombardment at the surface) or in capped structures, see Ref.~\cite{EbingerOxidationPRB1976} and Ref.~\cite{voltageAnnealing}, respectively.  In addition, other physical material systems, e.g. magnetic \cite{MagneticAnnealing} and piezo-electric \cite{PiezoElectricAnnealing}, can have their properties improved by applying an alternating field polarity \cite{ftj, ShajilNair2023}. While early work on the Al/Pb system, Ref.~\cite{APAA_Konkin_1982}, showed promising results on charge motion in- and out-of the junction, trapping and other modifications of AlO$_x$  using an alternating technique, no systematic approach to trimming the resistance and controllably modifying Al/AlO$_x$/Al junction morphology has been investigated and implemented.    

\section*{Results and Discussion}
In this work, we \change investigate the ABAA  technique, illustrated in Fig~\ref{Fig:CabreraMottVsABAA}(b) and Fig.~\ref{Fig:Bias_Schemes}, to modify the resistance of thin amorphous tunnel junctions. \stopchange    The ABAA technique entails applying an alternating polarity voltage pulse train to the \change as-grown \stopchange junctions during a global, low-temperature anneal. This is in sharp contrast to results obtained using only unipolar voltage pulses. For example, as shown in Fig.~\ref{Fig:Bias_Schemes}(d), when unipolar pulse trains were applied to the junctions a relatively small change in the resistance was observed. Changes of about 2\% and 6\%, respectively, were registered for negative-only and positive-only voltages after about 200 pulses. However, when alternating-polarity-bias (AB) pulses were used, as illustrated in Fig.~\ref{Fig:Bias_Schemes}(c), resistance changes of more than 70\% were recorded as shown in Fig.~\ref{Fig:Bias_Schemes}(d). In addition, we note two surprising features of the ABAA process that we address  below: first, that a cyclic response was typically observed for the AB pulse trains, as illustrated in the Fig.~\ref{Fig:Bias_Schemes}(d) inset; and second, a systematic jump in the resistance curve typically occurs during the process. 

Several features of Fig.~\ref{Fig:Bias_Schemes}(d) are instructive. For example, the asymmetry between the response to negative and positive pulse trains can be explained by considering the Cabrera-Mott mechanism for the oxidation. Since the initial oxidation involves only a negative bias, due to the charge transfer to the adsorbed oxygen, then relatively little change is expected when we re-apply that with the deposited top electrode. This can also create frustration, whereby cations and anions find themselves in metastable positions. When the voltage is reversed, anions and cations in the amorphous matrix can move more freely in the opposite direction, finding more stable positions, thereby allowing for a change in the resistance, as explained by the molecular dynamics simulations in Ref.~\cite{AlOxJunctionsLordiAPL2020}. 

Moreover, when the bias is sequentially alternated, the two species can then begin to migrate and diffuse, creating a mixing effect that apparently accentuates this effect, leading to an anomalously high increase in resistance. This process is analogous to, for example, degaussing a magnetized sample, indicating that the junctions are slightly polarized even after adding the top electrode. This predication is substantiated later in this work in the imaging section.

In addition, the cyclic, stepped response as the alternating process proceeds can be attributed to differential relaxation, not unlike \change in \stopchange high-field pulsed magnetic samples \cite{MagneticRelaxation} with spin-glass properties. 

\begin{figure}[hb]
\includegraphics[ width=\linewidth]{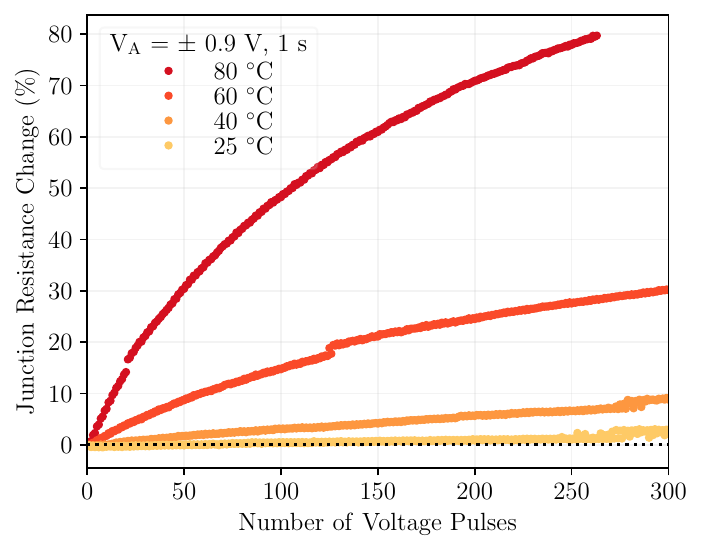}
\caption{\change\\Title: Junction resistance change as a function of the number of ABAA pulses\\
Legend: Data from four nominally identical junctions from the same wafer during the ABAA process at different substrate temperatures. In addition to the relatively smooth increase in junction resistance as the AB pulses are applied, discontinuities can often be observed as single events as in the process conducted at higher temperatures. In general, the annealing voltage amplitudes (for either polarity), pulse duration, and temperature are related.   The resistance tends to saturate after many pulses during the process, however, under extended cycling the junctions can occasionally break down, consistent with nano-mechanical fatigue. After the process is complete, a slight overshoot of the resistance is observed, consistent and comparable to that reported for laser-annealing in Ref.~\cite{LasiQBerkeley2022}. The uncertainty in the resistance measurements is less than the size of the points.\stopchange}
\label{Fig:ABAAvsTandA}
\end{figure}

This behaviour is consistent with the  discontinuities observed in Fig.~\ref{Fig:Bias_Schemes}(d) as well as in Fig.~\ref{Fig:ABAAvsTandA}  after fewer and fewer pulses (i.e. shorter times) as the temperature is increased. The discontinuities could be explained by the hypothesis that the effective temperature of the sample is higher due to the ABAA, and the junction undergoes a phase transition from amorphous to glass, as described in Ref.~\cite{Al2O3GlassTransitionTHashimoto2022} for T$\sim$400$\degree$. 
   
The temperature dependence of the ABAA from  Fig.~\ref{Fig:ABAAvsTandA} also shows that the rate of resistance enhancement increases significantly as the temperature goes up.    The increase is on the order of 2$\times$ every 10$\degree$C, consistent with that expected from an Arrhenius law. The ABAA effect was observed on all of the junctions in this study, with suppressed response observed from devices that had previously been heated above room temperature. Moreover, the size dependence (shown in \change Supplementary Note 1 and Figure 1\stopchange)  has relatively small correction over nearly an order of magnitude in area, indicating that the mechanism is likely not due to the perimeter or other edge effects such as diffusion and/or surface migration. Rather, this points to local effects such as reordering and increased coordination of the atoms inside the barrier. This can lead to changes in the effective barrier height as atoms reach a more stable configuration \cite{AlOxJunctionsLordiAPL2020}. Furthermore, since the data was taken at constant voltage, we infer that it is not a current-driven diffusion effect. These data \change hold \stopchange significant implications from the perspective of SQUID-based devices, such as sensors and tunable qubits, because these devices are typically made using two parallel junctions that may or may not be the same size.    

\begin{figure}[tp]
\includegraphics[, width=\linewidth]{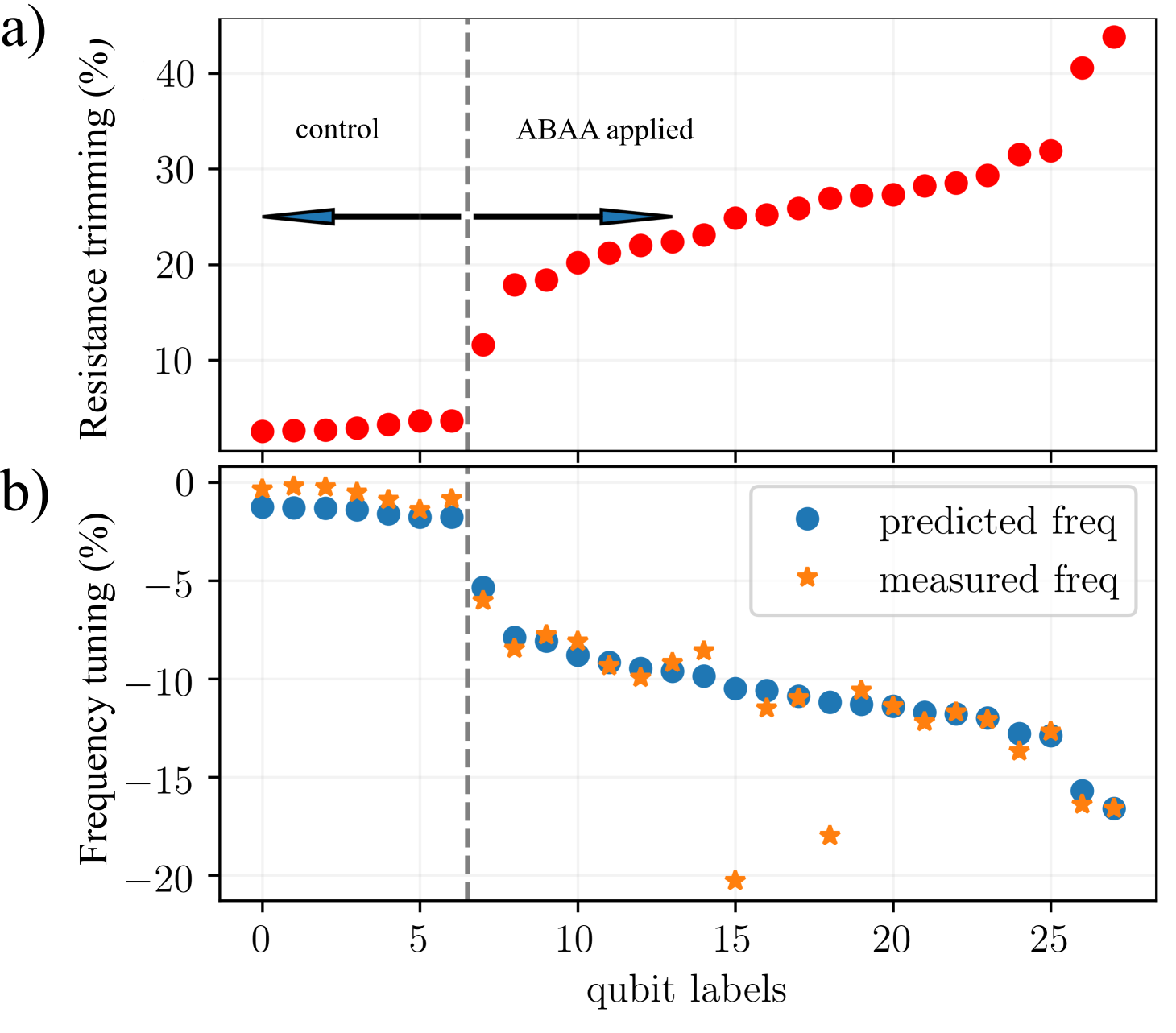}
\caption{\change\\Title: Translation of ABAA resistance trimming to frequency tuning. \\
Legend: Panel (a): Measured junction resistance shift (red dots) resulting from the ABAA trimming of junctions. Control devices without the ABAA applied were included to confirm that the changes were due to the ABAA process and calibrate the Ambegaokar-Baratoff parameters, and ABAA was applied to the rest of the junctions.\\ 
Panel (b): The blue dots show the predicted frequency tuning from the resistance before and after the process for both the control ABAA treated junctions. The orange stars show the actual, measured frequency tuning of each qubit relative to the predicted values. Prediction error is typically ~20MHz, giving error bars that are about the size of the points.\stopchange}
\label{Fig:Fig_QubitFrequencies}
\end{figure}

\begin{figure}[ht]
\includegraphics[, width=\columnwidth]{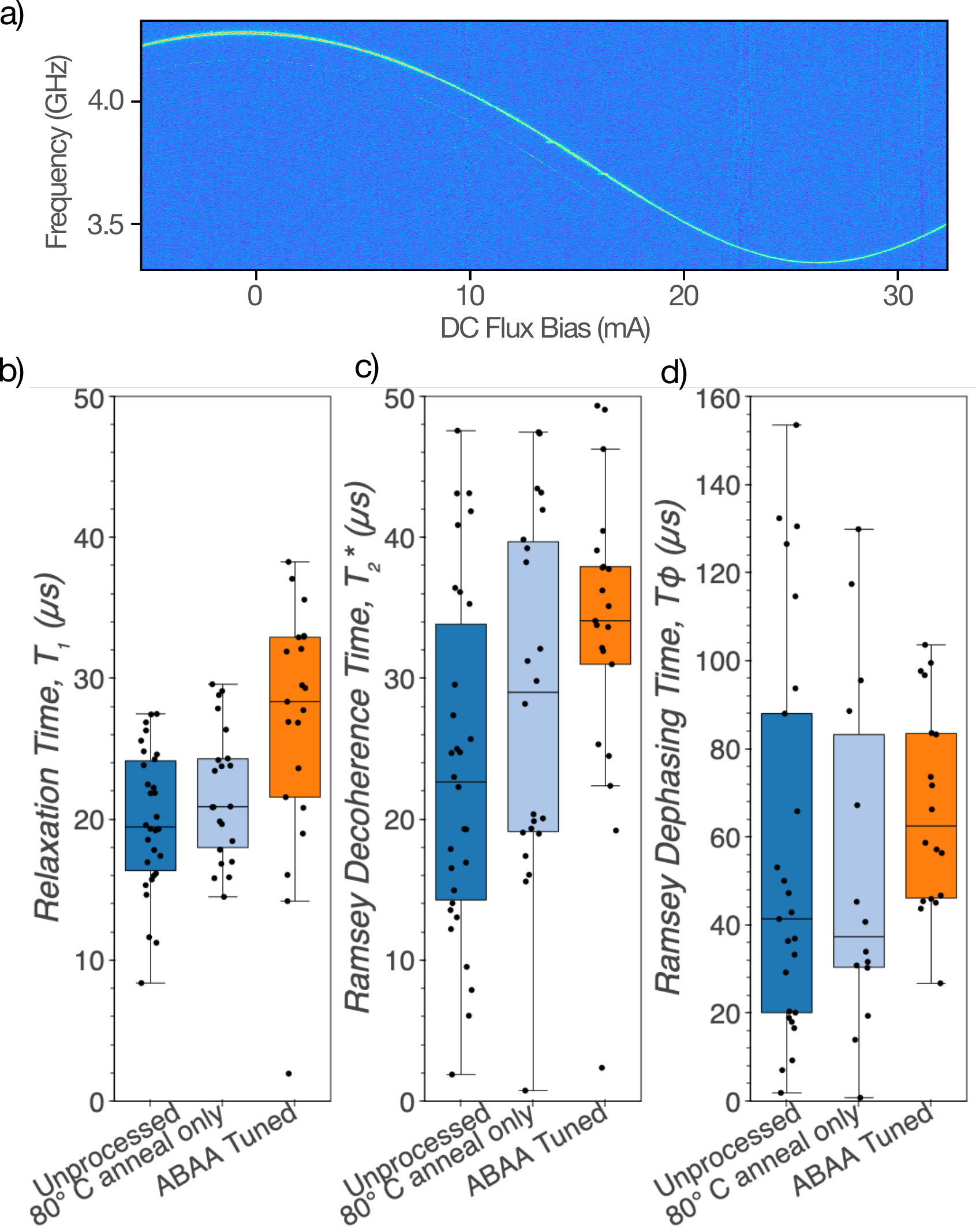}
\caption{\change\\Title: Spectroscopic and coherence properties of the qubits.\\
Panel (a): A representative
spectrum of a tunable qubit that has been treated with ABAA. The two
small features at approximately 3.8 GHz and 3.7 GHz
correspond to neighboring qubits.\\
Panels (b)-(d): Qubit coherence metrics $T_1$, $T_2^*$, and $T_{\phi}$, respectively. The study included two types of controls, i.e. multiple unprocessed qubits and qubits heated to 80 $\degree$C only. The ABAA tuned qubits were processed at $V_A=\pm0.8$~V, 1~s and T = $80\degree$C. The filled boxes error boxes span the 1st and 3rd quartile of the data uncertainty, while the whiskers mark the furthest datum within the $1.5\times$ the interquartile range past the 1st and 3rd quartiles.\stopchange}
\label{Fig:Fig_Coherence}
\end{figure}

We tested the effect of ABAA on the tunneling properties of the barrier by characterizing transmon qubits \change processed \stopchange with and without ABAA. We measured a variety of properties associated with junctions used in both tunable and fixed-frequency qubits~\cite{RigettiQubits}: whether $I_C$ follows the behavior expected by the Ambegaokar-Baratoff relationship, Fig~\ref{Fig:Fig_QubitFrequencies}; the presence of strongly coupled TLS's,\ref{Fig:Fig_Coherence}(a); and the impact of the barrier on qubit coherence, \ref{Fig:Fig_Coherence}(b).
 
The Ambegaokar-Baratoff parameters for this ensemble \change were \stopchange first determined using the untreated devices. Then, the measured frequencies of the treated qubits were calculated for various resistance-trimming amounts. For the non-tunable qubits, this is straightforward, while for the tunable qubits, it is necessary to find the maximum frequency, i.e. zero flux-bias, point using the qubit spectroscopy, shown in Fig.~\ref{Fig:Fig_Coherence}(a).
 
Figure ~\ref{Fig:Fig_QubitFrequencies}(a) shows the resistance change \change on several of the control qubits that were only \stopchange processed at 80$\degree$C and the 21 ABAA-processed qubits. The ABAA processed chips had typical junctions sizes, from $\sim$0.05 to $\sim$0.35 $\mu m^2$. The junctions are ordered by the amount that their resistance changed during the processing. \change Critical currents from 20   to   30 nA for the treated junctions were extracted along with the charging and Josephson energies of the qubits (see Supplementary Note 2 and Figure 2). \stopchange It can be seen in Figure~\ref{Fig:Fig_QubitFrequencies}(b) that the predicted and measured frequency tuning for ABAA-treated qubits match well, within 0.9$\pm$2.4\%, demonstrating that the trimmed resistances, as-measured after ABAA processing, can be used in the Ambegaokar-Baratoff formula. We find that this is consistent between junctions with ABAA applied vs. unprocessed with values of $(1.32\pm0.084)$~pH/$\Omega$ and $(1.27\pm0.109)$~pH/$\Omega$ respectively.   

The success rate of the ABAA-process for the initial tests on a process-test wafer was 79\%, \change as discussed in Supplementary Note 3.1. \stopchange In a subsequent extended study to optimize the ABAA process, a yield of 97.4$\pm0.4$\% was observed, \change see Supplementary Note 3.2. \stopchange

The qubit spectroscopy also probes for the existence of avoided level crossings in this frequency range.  Such crossings are signatures of strongly coupled TLS defects within the tunnel barrier   {\cite{JunctionVsSurfaceTLS_Lisenfeld2022,JunctionVsSurfaceTLS_Lisenfeld2019}.}  For reference, measurements on untreated qubits show about $0.7$ TLS/GHz, and on those that were heated to $80\degree$C without ABAA have $0.6$\,TLS/GHz.
Compared to these reference measurements, for the nine tunable devices in the ABAA-treated ensemble of this study with 5.4 GHz of tunable range, we saw no evidence of avoided level crossings larger than our $\sim$500 kHz resolution.  Figure~\ref{Fig:Fig_Coherence}(a) shows typical spectroscopy from an example qubit. Several other examples of TLS spectroscopy, both treated and not, from the ensemble are discussed in \change Supplementary Note 4, Figures 3-6\stopchange. We note here that smaller splittings, typically detected with swap spectroscopy \cite{JunctionVsSurfaceTLS_Lisenfeld2019} and likely due to surface defects, would not be affected by the ABAA. While this small data set is not conclusive evidence that such defects are suppressed by the ABAA process, it does point to the possibility that ABAA may improve qubit coherence and frequency stability. For example, the low-frequency stability of one of the devices was monitored over many hours (see \change Supplementary Note 5 and Figure 7\stopchange), and there was no evidence of coupling to TLFs \cite{SchlorDecoherencePRL}. In order to investigate this issue further, we studied the loss and time domain coherence in the ABAA qubit ensemble. 

\change The coherence metrics are shown in Figure~\ref{Fig:Fig_Coherence}(b)-(d) for all qubits measured in the initial study, and corresponding loss is discussed and shown in the Supplementary Note 6 and Table 1. \stopchange Here, each data point corresponds to the median of at least 60 measurements on each qubit in the ensemble of $\approx$ 70 qubits. (see discussion in \change Supplementary Note 3.1 \stopchange) We observe that the ABAA processed devices have a higher median $T_1$ than both the unprocessed   devices  and the devices that were only heated to $80\degree$C, although the $T_2^*$ and $T_\phi$ values between sample sets are not distinguishable within the spread of the values.
 
When analyzing processes that may reduce loss in qubits, it is useful to compare these metrics in the form of a qubit decay rate normalized to frequency  ($1/(T_1 \times (2 \pi f_{01})$); the ABAA tuned sample set has median loss tangent of $1.4\times10^{-6}$, while both the $80\degree$C annealed and unprocessed sets have median loss tangents of  $1.7\times 10^{-6}$.  We found that the coherence was typically improved after the ABAA process, indicating that the resistance trimming may have reduced the median loss tangent in the   qubits    by about $0.3\times 10^{-6}$.   These qubits are similar to devices in Ref.~\cite{nersisyan2019manufacturing} and "Design B" from Ref.~\cite{wang2015surface}, where the junction has relatively low participation. Hence, the modest increase in $T_1$ reflects the electrode surfaces/materials-centric design of these qubits.     

\begin{figure}[ht]
\includegraphics[, width=\linewidth]{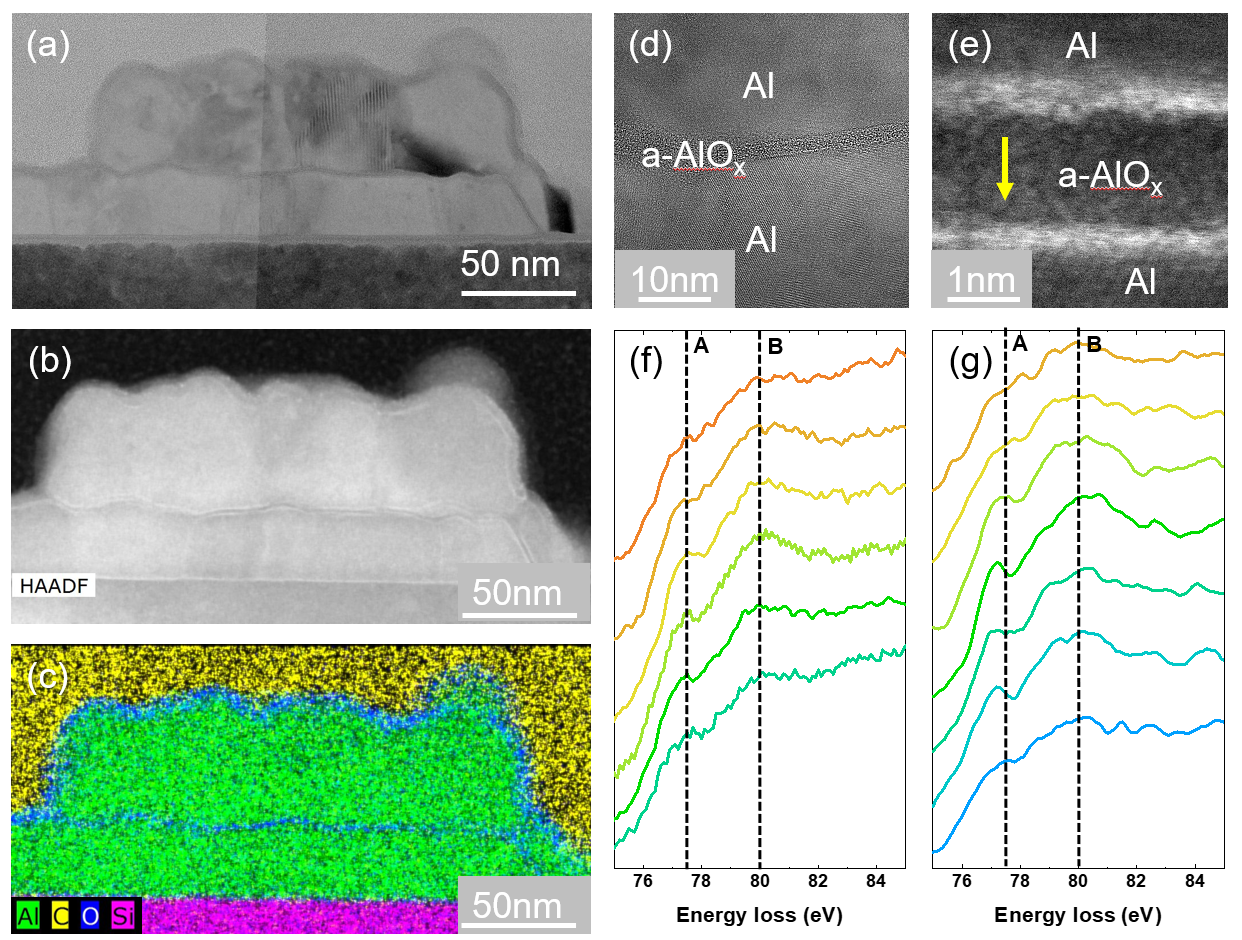}
\caption{\change \\Title: Electron microscopy analysis on ABAA-treated junction.\\
Legend: Panel (a): Low magnification cross-sectional TEM image of a junction.\\
Panel (b): HAADF-STEM image of the same junction.\\
Panel (c): Corresponding STEM-EDS elemental distribution map of the junction.\\ 
Panel (d) High-resolution TEM image of the Al/a-AlO$_x$/Al interfaces.\\ 
Panel (e): HAADF-STEM image of Al/a-AlO$_x$/Al interfaces.\\ 
Panel (f): ELNES of Al23 edge of the ABAA-treated sample taken from the yellow arrow indicated region in Panel (e)\\
Panel (g): ELNES of Al23 edge of the untreated a-AlO$_x$ barrier from reference \cite{ALD_Oh_2024}\stopchange}
\label{Fig:Fig_TEM}
\end{figure}
\change Finally, transmission electron microscopy (TEM) imaging and scanning TEM (STEM) with electron energy-loss near-edge structure (ELNES) analysis and high-angle-annual dark-field (HAADF) imaging, respectively were employed.\stopchange These allowed us  to investigate morphological changes in the junctions due to the ABAA process. Details for this study are provided in \change Supplementary Note 7\stopchange. These studies indicate that the increase in junction resistivity may be attributed to a more uniform barrier structure with reduced point defect density.  A low-mag cross-sectional TEM/STEM image and corresponding EDS elemental distribution map are shown in Fig.~\ref{Fig:Fig_TEM}(a-c) showing that a-AlO$_x$ barrier is still amorphous in the ABAA-treated sample (Fig.~\ref{Fig:Fig_TEM}(d)), and the HAADF-STEM image shows a barrier thickness of approximately 2 nm (Fig. ~\ref{Fig:Fig_TEM}(e)). 

While the ABAA technique does not appear to change the amorphous nature of the barrier from visual inspection, the chemical analysis tells a different story. We see a much more homogeneous chemistry in the ABAA-treated a-AlO$_x$ than in untreated junctions. Our previous research showed that the distance between two main peaks labeled as A and B in Fig.~\ref{Fig:Fig_TEM}(f-g), close to approximately 77.5 eV and 80 eV of the Al L23 energy-loss near edge structure (ELNES) respectively, can be used to analyze the coordination number of Al \cite{ALD_Oh_2024, BouchetAluminaUltraMicroscopy2003,FritzAlOxNanostructureScientificReports2018,FritzStructureMechancialAlOxPhysRevMater2019, KimotoCoordinationAl203APL2003,WeigelHRTEMJofPhysCondMat2008}. Specifically, the Al L3 edges (peak A) tends to shift to higher energy as Al coordination numbers increase. For untreated junctions, a spatial variation in coordination number across the a-AlO$_x$ barrier from the untreated junction (Fig.~\ref{Fig:Fig_TEM}(g)) \cite{ALD_Oh_2024} is demonstrated and considered as a potential decoherence source. In contrast, the ABAA-treated junction shows a uniform distribution of Al coordination number through the barrier, as the peak positions of A and B remain unchanged across the a-AlO$_x$ layer, as shown in Fig.~\ref{Fig:Fig_TEM}(f). This may originate from the migration/recombination of charged point defects induced by the alternating bias field \cite{DislocationMotionLiNatureMaterials2023} giving rise to a more chemically homogeneous barrier layer. This effect may increase the effective barrier height of the tunnel junction leading to the observed increase in resistance.

 \section*{Conclusion}
We find that an alternating bias is effective in assisting the thermal annealing of   ionically-bonded    amorphous materials.   The effect is attributed to depolarization of the as-fabricated amorphous matrix as cations and anions are driven in opposite directions, allowing them to find more stable positions,    similar to adding restarts in simulated annealing solvers to avoid local minima \cite{SimulatedAnnealingResetAlfonzetti2006}. This allows for significant modifications of the junction resistance, in excess of +70$\%$, with apparent reductions in loss and defects.  Junctions that were tested, however, we do not  These modifications can be made easily with a standard probe station at the wafer, die, or device level. In practice, we find the exponential increase is stable and tends to saturate at a high number of pulses. The behavior of the resistance during the ABAA process showed cyclic small and large changes in resistance as the positive- and negative-bias voltages were applied and the junction resistance increased, and typically a single discontinuity in resistance was observed if a sufficient number of pulses were applied. 

These results along with the temperature-dependent studies and imaging point to an assisted annealing mechanism, for example, improved local ordering and/or the a-AlO$_x$ being driven into a lower-energy glassy state \cite{Al2O3GlassTransitionTHashimoto2022,GlassTransitionDebenedettiNature2001, GlassyJunctionsPhysRevB} in the potential energy landscape, as is indicated by the observed reduction in loss and TLS. In order to confirm these hypotheses, more studies such as high-resolution imaging and molecular dynamics that include density-functional theory are required. Regardless of the mechanism, we expect that a controlled alternating bias-assisted junction annealing approach will pave the way for building large-scale superconducting quantum processors by improving qubit frequency targeting, coherence, and stability. Other applications where targeting tunnel junction resistance is important may benefit by allowing for more homogeneous junction chains and ensembles for devices ranging from amplifiers to voltage standards, sensors, and data storage. 

\change
\section*{Methods}
\stopchange
The ABAA treatment was implemented on as-grown junctions using a conventional probe station with a low-temperature heating stage, as illustrated in Fig.~\ref{Fig:Bias_Schemes}~(a). Using this technique, it can be expected that the Al$^{3+}$ cations and O$^{2-}$ anions in the barrier will be driven in opposite directions at each pulse. By alternating the bias voltage the ions can be driven out of local, metastable sites into a lower energy configuration. In general, the guidelines used for the voltage profiles were that it should be on the order of $V$\textsubscript{Mott} and less than the breakdown voltage, $V_B\sim1.5$~V \cite{AlOxBreakdownJAP2004,TemperatureAndPressureMottPotentials2012}  , as discussed and illustrated in Supplementary Note 8 and Figure 9, respectively.   When considering the time duration of the pulse, we note that by staying well below $V_B$ the junction properties were observed to evolve relatively slowly, on the order of   hundreds    of seconds, similar to the known oxidation rates of these surfaces. Finally, a relatively low temperature is desired to not expose junctions to a higher temperature than the highest temperature they are exposed to during the fabrication process. 

The equipment and pulse sequences used for the tuning consist of a standard source-measure unit (SMU) and a probe station with a heated platen, illustrated in Fig.~\ref{Fig:Bias_Schemes}(a,b). The SMU was programmed to provide a train of 1-second-long anneal-assist pulses after the wafer was brought to the target temperature. The pulse amplitudes were set in the range of 0.8 - 1.1 V. The resistance of the junctions was measured after each pulse with an additional pulse at low voltage. 

\change
\section*{Data availability}
The datasets generated during and/or analysed during the current study are available from the corresponding authors on reasonable request.
\stopchange

\onecolumngrid{ \section*{References}}
\twocolumngrid
\bibliography{VIA.bib}

\providecommand{\noopsort}[1]{}\providecommand{\singleletter}[1]{#1}%
\begin{thebibliography}{58}%
\makeatletter
\providecommand \@ifxundefined [1]{%
 \@ifx{#1\undefined}
}%
\providecommand \@ifnum [1]{%
 \ifnum #1\expandafter \@firstoftwo
 \else \expandafter \@secondoftwo
 \fi
}%
\providecommand \@ifx [1]{%
 \ifx #1\expandafter \@firstoftwo
 \else \expandafter \@secondoftwo
 \fi
}%
\providecommand \natexlab [1]{#1}%
\providecommand \enquote  [1]{``#1''}%
\providecommand \bibnamefont  [1]{#1}%
\providecommand \bibfnamefont [1]{#1}%
\providecommand \citenamefont [1]{#1}%
\providecommand \href@noop [0]{\@secondoftwo}%
\providecommand \href [0]{\begingroup \@sanitize@url \@href}%
\providecommand \@href[1]{\@@startlink{#1}\@@href}%
\providecommand \@@href[1]{\endgroup#1\@@endlink}%
\providecommand \@sanitize@url [0]{\catcode `\\12\catcode `\$12\catcode `\&12\catcode `\#12\catcode `\^12\catcode `\_12\catcode `\%12\relax}%
\providecommand \@@startlink[1]{}%
\providecommand \@@endlink[0]{}%
\providecommand \url  [0]{\begingroup\@sanitize@url \@url }%
\providecommand \@url [1]{\endgroup\@href {#1}{\urlprefix }}%
\providecommand \urlprefix  [0]{URL }%
\providecommand \Eprint [0]{\href }%
\providecommand \doibase [0]{https://doi.org/}%
\providecommand \selectlanguage [0]{\@gobble}%
\providecommand \bibinfo  [0]{\@secondoftwo}%
\providecommand \bibfield  [0]{\@secondoftwo}%
\providecommand \translation [1]{[#1]}%
\providecommand \BibitemOpen [0]{}%
\providecommand \bibitemStop [0]{}%
\providecommand \bibitemNoStop [0]{.\EOS\space}%
\providecommand \EOS [0]{\spacefactor3000\relax}%
\providecommand \BibitemShut  [1]{\csname bibitem#1\endcsname}%
\let\auto@bib@innerbib\@empty
\bibitem [{\citenamefont {Josephson}(1962)}]{JOSEPHSON1962PRL}%
  \BibitemOpen
  \bibfield  {author} {\bibinfo {author} {\bibfnamefont {B.}~\bibnamefont {Josephson}},\ }\bibfield  {title} {\bibinfo {title} {Possible new effects in superconductive tunnelling},\ }\href {https://doi.org/https://doi.org/10.1016/0031-9163(62)91369-0} {\bibfield  {journal} {\bibinfo  {journal} {Physics Letters}\ }\textbf {\bibinfo {volume} {1}},\ \bibinfo {pages} {251} (\bibinfo {year} {1962})}\BibitemShut {NoStop}%
\bibitem [{\citenamefont {Clarke}(2004)}]{Clarke2004}%
  \BibitemOpen
  \bibfield  {author} {\bibinfo {author} {\bibfnamefont {J.}~\bibnamefont {Clarke}},\ }\href {http://inis.iaea.org/search/search.aspx?orig_q=RN:38047859} {\emph {\bibinfo {title} {The SQUID handbook Vol 1 Fundamentals and technology of SQUIDs and SQUID systems}}}\ (\bibinfo  {publisher} {Wiley VCH},\ \bibinfo {address} {Germany},\ \bibinfo {year} {2004})\BibitemShut {NoStop}%
\bibitem [{\citenamefont {Devoret}(1997)}]{Devoret1997}%
  \BibitemOpen
  \bibfield  {author} {\bibinfo {author} {\bibfnamefont {M.~H.}\ \bibnamefont {Devoret}},\ }\href {http://inis.iaea.org/search/search.aspx?orig_q=RN:29063476} {\emph {\bibinfo {title} {Quantum fluctuations in electrical circuits}}}\ (\bibinfo  {publisher} {Edition de Physique},\ \bibinfo {address} {France},\ \bibinfo {year} {1997})\BibitemShut {NoStop}%
\bibitem [{\citenamefont {Nakamura}\ \emph {et~al.}(1999)\citenamefont {Nakamura}, \citenamefont {Pashkin},\ and\ \citenamefont {Tsai}}]{Nakamura1999}%
  \BibitemOpen
  \bibfield  {author} {\bibinfo {author} {\bibfnamefont {Y.}~\bibnamefont {Nakamura}}, \bibinfo {author} {\bibfnamefont {Y.~A.}\ \bibnamefont {Pashkin}},\ and\ \bibinfo {author} {\bibfnamefont {J.~S.}\ \bibnamefont {Tsai}},\ }\bibfield  {title} {\bibinfo {title} {Coherent control of macroscopic quantum states in a single-cooper-pair box},\ }\href {https://doi.org/10.1038/19718} {\bibfield  {journal} {\bibinfo  {journal} {Nature}\ }\textbf {\bibinfo {volume} {398}},\ \bibinfo {pages} {786} (\bibinfo {year} {1999})}\BibitemShut {NoStop}%
\bibitem [{\citenamefont {Castellanos-Beltran}\ \emph {et~al.}(2009)\citenamefont {Castellanos-Beltran}, \citenamefont {Irwin}, \citenamefont {Vale}, \citenamefont {Hilton},\ and\ \citenamefont {Lehnert}}]{JPA2009}%
  \BibitemOpen
  \bibfield  {author} {\bibinfo {author} {\bibfnamefont {M.~A.}\ \bibnamefont {Castellanos-Beltran}}, \bibinfo {author} {\bibfnamefont {K.~D.}\ \bibnamefont {Irwin}}, \bibinfo {author} {\bibfnamefont {L.~R.}\ \bibnamefont {Vale}}, \bibinfo {author} {\bibfnamefont {G.~C.}\ \bibnamefont {Hilton}},\ and\ \bibinfo {author} {\bibfnamefont {K.~W.}\ \bibnamefont {Lehnert}},\ }\bibfield  {title} {\bibinfo {title} {Bandwidth and dynamic range of a widely tunable josephson parametric amplifier},\ }\href {https://doi.org/10.1109/TASC.2009.2018119} {\bibfield  {journal} {\bibinfo  {journal} {IEEE Transactions on Applied Superconductivity}\ }\textbf {\bibinfo {volume} {19}},\ \bibinfo {pages} {944} (\bibinfo {year} {2009})}\BibitemShut {NoStop}%
\bibitem [{\citenamefont {Macklin}\ \emph {et~al.}(2015)\citenamefont {Macklin}, \citenamefont {O’Brien}, \citenamefont {Hover}, \citenamefont {Schwartz}, \citenamefont {Bolkhovsky}, \citenamefont {Zhang}, \citenamefont {Oliver},\ and\ \citenamefont {Siddiqi}}]{TWPA2015}%
  \BibitemOpen
  \bibfield  {author} {\bibinfo {author} {\bibfnamefont {C.}~\bibnamefont {Macklin}}, \bibinfo {author} {\bibfnamefont {K.}~\bibnamefont {O’Brien}}, \bibinfo {author} {\bibfnamefont {D.}~\bibnamefont {Hover}}, \bibinfo {author} {\bibfnamefont {M.~E.}\ \bibnamefont {Schwartz}}, \bibinfo {author} {\bibfnamefont {V.}~\bibnamefont {Bolkhovsky}}, \bibinfo {author} {\bibfnamefont {X.}~\bibnamefont {Zhang}}, \bibinfo {author} {\bibfnamefont {W.~D.}\ \bibnamefont {Oliver}},\ and\ \bibinfo {author} {\bibfnamefont {I.}~\bibnamefont {Siddiqi}},\ }\bibfield  {title} {\bibinfo {title} {A near–quantum-limited josephson traveling-wave parametric amplifier},\ }\href {https://doi.org/10.1126/science.aaa8525} {\bibfield  {journal} {\bibinfo  {journal} {Science}\ }\textbf {\bibinfo {volume} {350}},\ \bibinfo {pages} {307} (\bibinfo {year} {2015})},\ \Eprint {https://arxiv.org/abs/https://www.science.org/doi/pdf/10.1126/science.aaa8525} {https://www.science.org/doi/pdf/10.1126/science.aaa8525} \BibitemShut {NoStop}%
\bibitem [{\citenamefont {Likharev}\ and\ \citenamefont {Semenov}(1991)}]{RSFQ1991}%
  \BibitemOpen
  \bibfield  {author} {\bibinfo {author} {\bibfnamefont {K.}~\bibnamefont {Likharev}}\ and\ \bibinfo {author} {\bibfnamefont {V.}~\bibnamefont {Semenov}},\ }\bibfield  {title} {\bibinfo {title} {Rsfq logic/memory family: a new josephson-junction technology for sub-terahertz-clock-frequency digital systems},\ }\href {https://doi.org/10.1109/77.80745} {\bibfield  {journal} {\bibinfo  {journal} {IEEE Transactions on Applied Superconductivity}\ }\textbf {\bibinfo {volume} {1}},\ \bibinfo {pages} {3} (\bibinfo {year} {1991})}\BibitemShut {NoStop}%
\bibitem [{\citenamefont {{Shajil Nair}}\ \emph {et~al.}(2023)\citenamefont {{Shajil Nair}}, \citenamefont {Holzer}, \citenamefont {Dubourdieu},\ and\ \citenamefont {Deshpande}}]{ShajilNair2023}%
  \BibitemOpen
  \bibfield  {author} {\bibinfo {author} {\bibfnamefont {K.}~\bibnamefont {{Shajil Nair}}}, \bibinfo {author} {\bibfnamefont {M.}~\bibnamefont {Holzer}}, \bibinfo {author} {\bibfnamefont {C.}~\bibnamefont {Dubourdieu}},\ and\ \bibinfo {author} {\bibfnamefont {V.}~\bibnamefont {Deshpande}},\ }\bibfield  {title} {\bibinfo {title} {{Cycling Waveform Dependent Wake-Up and ON/OFF Ratio in Al2O3/Hf0.5Zr0.5O2 Ferroelectric Tunnel Junction Devices}},\ }\href {https://doi.org/10.1021/acsaelm.2c01492} {\bibfield  {journal} {\bibinfo  {journal} {ACS Applied Electronic Materials}\ }\textbf {\bibinfo {volume} {5}},\ \bibinfo {pages} {1478} (\bibinfo {year} {2023})}\BibitemShut {NoStop}%
\bibitem [{\citenamefont {Zhu}\ and\ \citenamefont {Park}(2006)}]{ZHU2006}%
  \BibitemOpen
  \bibfield  {author} {\bibinfo {author} {\bibfnamefont {J.-G.~J.}\ \bibnamefont {Zhu}}\ and\ \bibinfo {author} {\bibfnamefont {C.}~\bibnamefont {Park}},\ }\bibfield  {title} {\bibinfo {title} {Magnetic tunnel junctions},\ }\href {https://doi.org/https://doi.org/10.1016/S1369-7021(06)71693-5} {\bibfield  {journal} {\bibinfo  {journal} {Materials Today}\ }\textbf {\bibinfo {volume} {9}},\ \bibinfo {pages} {36} (\bibinfo {year} {2006})}\BibitemShut {NoStop}%
\bibitem [{\citenamefont {Rippard}\ \emph {et~al.}(2002)\citenamefont {Rippard}, \citenamefont {Perrella}, \citenamefont {Albert},\ and\ \citenamefont {Buhrman}}]{BEEM_RippardPRL2002}%
  \BibitemOpen
  \bibfield  {author} {\bibinfo {author} {\bibfnamefont {W.~H.}\ \bibnamefont {Rippard}}, \bibinfo {author} {\bibfnamefont {A.~C.}\ \bibnamefont {Perrella}}, \bibinfo {author} {\bibfnamefont {F.~J.}\ \bibnamefont {Albert}},\ and\ \bibinfo {author} {\bibfnamefont {R.~A.}\ \bibnamefont {Buhrman}},\ }\bibfield  {title} {\bibinfo {title} {Ultrathin aluminum oxide tunnel barriers},\ }\href {https://doi.org/10.1103/PhysRevLett.88.046805} {\bibfield  {journal} {\bibinfo  {journal} {Phys. Rev. Lett.}\ }\textbf {\bibinfo {volume} {88}},\ \bibinfo {pages} {046805} (\bibinfo {year} {2002})}\BibitemShut {NoStop}%
\bibitem [{\citenamefont {O'Dwyer}(2004)}]{AlOxBreakdownJAP2004}%
  \BibitemOpen
  \bibfield  {author} {\bibinfo {author} {\bibfnamefont {J.~J.}\ \bibnamefont {O'Dwyer}},\ }\bibfield  {title} {\bibinfo {title} {{Current‐Voltage Characteristics of Dielectric Films}},\ }\href {https://doi.org/10.1063/1.1708222} {\bibfield  {journal} {\bibinfo  {journal} {Journal of Applied Physics}\ }\textbf {\bibinfo {volume} {37}},\ \bibinfo {pages} {599} (\bibinfo {year} {2004})},\ \Eprint {https://arxiv.org/abs/https://pubs.aip.org/aip/jap/article-pdf/37/2/599/7938519/599\_1\_online.pdf} {https://pubs.aip.org/aip/jap/article-pdf/37/2/599/7938519/599\_1\_online.pdf} \BibitemShut {NoStop}%
\bibitem [{\citenamefont {Koch}\ \emph {et~al.}(2007)\citenamefont {Koch}, \citenamefont {Yu}, \citenamefont {Gambetta}, \citenamefont {Houck}, \citenamefont {Schuster}, \citenamefont {Majer}, \citenamefont {Blais}, \citenamefont {Devoret}, \citenamefont {Girvin},\ and\ \citenamefont {Schoelkopf}}]{koch2007charge}%
  \BibitemOpen
  \bibfield  {author} {\bibinfo {author} {\bibfnamefont {J.}~\bibnamefont {Koch}}, \bibinfo {author} {\bibfnamefont {T.}~\bibnamefont {Yu}}, \bibinfo {author} {\bibfnamefont {J.}~\bibnamefont {Gambetta}}, \bibinfo {author} {\bibfnamefont {A.}~\bibnamefont {Houck}}, \bibinfo {author} {\bibfnamefont {D.}~\bibnamefont {Schuster}}, \bibinfo {author} {\bibfnamefont {J.}~\bibnamefont {Majer}}, \bibinfo {author} {\bibfnamefont {A.}~\bibnamefont {Blais}}, \bibinfo {author} {\bibfnamefont {M.}~\bibnamefont {Devoret}}, \bibinfo {author} {\bibfnamefont {S.}~\bibnamefont {Girvin}},\ and\ \bibinfo {author} {\bibfnamefont {R.}~\bibnamefont {Schoelkopf}},\ }\bibfield  {title} {\bibinfo {title} {Charge-insensitive qubit design derived from the cooper pair box},\ }\href@noop {} {\bibfield  {journal} {\bibinfo  {journal} {Phys. Rev. A}\ }\textbf {\bibinfo {volume} {76}},\ \bibinfo {pages} {042319} (\bibinfo {year} {2007})}\BibitemShut {NoStop}%
\bibitem [{\citenamefont {van Duzer}(1999)}]{VanDuzer}%
  \BibitemOpen
  \bibfield  {author} {\bibinfo {author} {\bibfnamefont {C.}~\bibnamefont {van Duzer}, \bibfnamefont {Theodore;~Turner}},\ }\href@noop {} {\emph {\bibinfo {title} {Principles of Superconductive Devices and Circuits, 2nd ed.}}}\ (\bibinfo  {publisher} {Prentice-Hall},\ \bibinfo {address} {Upper Saddle River NJ},\ \bibinfo {year} {1999})\BibitemShut {NoStop}%
\bibitem [{\citenamefont {Muthusubramanian}\ \emph {et~al.}(2023)\citenamefont {Muthusubramanian}, \citenamefont {Finkel}, \citenamefont {Duivestein}, \citenamefont {Zachariadis}, \citenamefont {van~der Meer}, \citenamefont {Veen}, \citenamefont {Beekman}, \citenamefont {Stavenga}, \citenamefont {Bruno},\ and\ \citenamefont {DiCarlo}}]{Muthusubramanian2023}%
  \BibitemOpen
  \bibfield  {author} {\bibinfo {author} {\bibfnamefont {N.}~\bibnamefont {Muthusubramanian}}, \bibinfo {author} {\bibfnamefont {M.}~\bibnamefont {Finkel}}, \bibinfo {author} {\bibfnamefont {P.}~\bibnamefont {Duivestein}}, \bibinfo {author} {\bibfnamefont {C.}~\bibnamefont {Zachariadis}}, \bibinfo {author} {\bibfnamefont {S.~L.~M.}\ \bibnamefont {van~der Meer}}, \bibinfo {author} {\bibfnamefont {H.~M.}\ \bibnamefont {Veen}}, \bibinfo {author} {\bibfnamefont {M.~W.}\ \bibnamefont {Beekman}}, \bibinfo {author} {\bibfnamefont {T.}~\bibnamefont {Stavenga}}, \bibinfo {author} {\bibfnamefont {A.}~\bibnamefont {Bruno}},\ and\ \bibinfo {author} {\bibfnamefont {L.}~\bibnamefont {DiCarlo}},\ }\bibfield  {title} {\bibinfo {title} {Wafer-scale uniformity of dolan-bridge and bridgeless manhattan-style josephson junctions for superconducting quantum processors},\ }\bibfield  {journal} {\bibinfo  {journal} {Quantum Science and Technology}\ }\href {https://doi.org/10.1088/2058-9565/ad199c} {10.1088/2058-9565/ad199c}
  (\bibinfo {year} {2023})\BibitemShut {NoStop}%
\bibitem [{\citenamefont {Zeng}\ \emph {et~al.}(2016)\citenamefont {Zeng}, \citenamefont {Tran}, \citenamefont {Tai}, \citenamefont {Svensson},\ and\ \citenamefont {Olsson}}]{Zeng2016}%
  \BibitemOpen
  \bibfield  {author} {\bibinfo {author} {\bibfnamefont {L.}~\bibnamefont {Zeng}}, \bibinfo {author} {\bibfnamefont {D.~T.}\ \bibnamefont {Tran}}, \bibinfo {author} {\bibfnamefont {C.-W.}\ \bibnamefont {Tai}}, \bibinfo {author} {\bibfnamefont {G.}~\bibnamefont {Svensson}},\ and\ \bibinfo {author} {\bibfnamefont {E.}~\bibnamefont {Olsson}},\ }\bibfield  {title} {\bibinfo {title} {{Atomic structure and oxygen deficiency of the ultrathin aluminium oxide barrier in Al/AlOx/Al Josephson junctions}},\ }\href {https://doi.org/10.1038/srep29679} {\bibfield  {journal} {\bibinfo  {journal} {Scientific Reports}\ }\textbf {\bibinfo {volume} {6}},\ \bibinfo {pages} {29679} (\bibinfo {year} {2016})}\BibitemShut {NoStop}%
\bibitem [{\citenamefont {Bilmes}\ \emph {et~al.}(2021)\citenamefont {Bilmes}, \citenamefont {Händel}, \citenamefont {Volosheniuk}, \citenamefont {Ustinov},\ and\ \citenamefont {Lisenfeld}}]{InsituBandaid_Bilmes_2021}%
  \BibitemOpen
  \bibfield  {author} {\bibinfo {author} {\bibfnamefont {A.}~\bibnamefont {Bilmes}}, \bibinfo {author} {\bibfnamefont {A.~K.}\ \bibnamefont {Händel}}, \bibinfo {author} {\bibfnamefont {S.}~\bibnamefont {Volosheniuk}}, \bibinfo {author} {\bibfnamefont {A.~V.}\ \bibnamefont {Ustinov}},\ and\ \bibinfo {author} {\bibfnamefont {J.}~\bibnamefont {Lisenfeld}},\ }\bibfield  {title} {\bibinfo {title} {In-situ bandaged josephson junctions for superconducting quantum processors},\ }\href {https://doi.org/10.1088/1361-6668/ac2a6d} {\bibfield  {journal} {\bibinfo  {journal} {Superconductor Science and Technology}\ }\textbf {\bibinfo {volume} {34}},\ \bibinfo {pages} {125011} (\bibinfo {year} {2021})}\BibitemShut {NoStop}%
\bibitem [{\citenamefont {Martinis}\ \emph {et~al.}(2005)\citenamefont {Martinis}, \citenamefont {Cooper}, \citenamefont {McDermott}, \citenamefont {Steffen}, \citenamefont {Ansmann}, \citenamefont {Osborn}, \citenamefont {Cicak}, \citenamefont {Oh}, \citenamefont {Pappas}, \citenamefont {Simmonds} \emph {et~al.}}]{martinis2005decoherence}%
  \BibitemOpen
  \bibfield  {author} {\bibinfo {author} {\bibfnamefont {J.}~\bibnamefont {Martinis}}, \bibinfo {author} {\bibfnamefont {K.}~\bibnamefont {Cooper}}, \bibinfo {author} {\bibfnamefont {R.}~\bibnamefont {McDermott}}, \bibinfo {author} {\bibfnamefont {M.}~\bibnamefont {Steffen}}, \bibinfo {author} {\bibfnamefont {M.}~\bibnamefont {Ansmann}}, \bibinfo {author} {\bibfnamefont {K.}~\bibnamefont {Osborn}}, \bibinfo {author} {\bibfnamefont {K.}~\bibnamefont {Cicak}}, \bibinfo {author} {\bibfnamefont {S.}~\bibnamefont {Oh}}, \bibinfo {author} {\bibfnamefont {D.}~\bibnamefont {Pappas}}, \bibinfo {author} {\bibfnamefont {R.}~\bibnamefont {Simmonds}}, \emph {et~al.},\ }\bibfield  {title} {\bibinfo {title} {Decoherence in josephson qubits from dielectric loss},\ }\href@noop {} {\bibfield  {journal} {\bibinfo  {journal} {Phys. Rev. Lett.}\ }\textbf {\bibinfo {volume} {95}},\ \bibinfo {pages} {210503} (\bibinfo {year} {2005})}\BibitemShut {NoStop}%
\bibitem [{\citenamefont {Schl\"or}\ \emph {et~al.}(2019)\citenamefont {Schl\"or}, \citenamefont {Lisenfeld}, \citenamefont {M\"uller}, \citenamefont {Bilmes}, \citenamefont {Schneider}, \citenamefont {Pappas}, \citenamefont {Ustinov},\ and\ \citenamefont {Weides}}]{SchlorDecoherencePRL}%
  \BibitemOpen
  \bibfield  {author} {\bibinfo {author} {\bibfnamefont {S.}~\bibnamefont {Schl\"or}}, \bibinfo {author} {\bibfnamefont {J.}~\bibnamefont {Lisenfeld}}, \bibinfo {author} {\bibfnamefont {C.}~\bibnamefont {M\"uller}}, \bibinfo {author} {\bibfnamefont {A.}~\bibnamefont {Bilmes}}, \bibinfo {author} {\bibfnamefont {A.}~\bibnamefont {Schneider}}, \bibinfo {author} {\bibfnamefont {D.~P.}\ \bibnamefont {Pappas}}, \bibinfo {author} {\bibfnamefont {A.~V.}\ \bibnamefont {Ustinov}},\ and\ \bibinfo {author} {\bibfnamefont {M.}~\bibnamefont {Weides}},\ }\bibfield  {title} {\bibinfo {title} {Correlating decoherence in transmon qubits: Low frequency noise by single fluctuators},\ }\href {https://doi.org/10.1103/PhysRevLett.123.190502} {\bibfield  {journal} {\bibinfo  {journal} {Phys. Rev. Lett.}\ }\textbf {\bibinfo {volume} {123}},\ \bibinfo {pages} {190502} (\bibinfo {year} {2019})}\BibitemShut {NoStop}%
\bibitem [{\citenamefont {Lecocq}\ \emph {et~al.}(2011)\citenamefont {Lecocq}, \citenamefont {Pop}, \citenamefont {Peng}, \citenamefont {Matei}, \citenamefont {Crozes}, \citenamefont {Fournier}, \citenamefont {Naud}, \citenamefont {Guichard},\ and\ \citenamefont {Buisson}}]{Lecocq2011}%
  \BibitemOpen
  \bibfield  {author} {\bibinfo {author} {\bibfnamefont {F.}~\bibnamefont {Lecocq}}, \bibinfo {author} {\bibfnamefont {I.~M.}\ \bibnamefont {Pop}}, \bibinfo {author} {\bibfnamefont {Z.}~\bibnamefont {Peng}}, \bibinfo {author} {\bibfnamefont {I.}~\bibnamefont {Matei}}, \bibinfo {author} {\bibfnamefont {T.}~\bibnamefont {Crozes}}, \bibinfo {author} {\bibfnamefont {T.}~\bibnamefont {Fournier}}, \bibinfo {author} {\bibfnamefont {C.}~\bibnamefont {Naud}}, \bibinfo {author} {\bibfnamefont {W.}~\bibnamefont {Guichard}},\ and\ \bibinfo {author} {\bibfnamefont {O.}~\bibnamefont {Buisson}},\ }\bibfield  {title} {\bibinfo {title} {Junction fabrication by shadow evaporation without a suspended bridge},\ }\href {https://doi.org/10.1088/0957-4484/22/31/315302} {\bibfield  {journal} {\bibinfo  {journal} {Nanotechnology}\ }\textbf {\bibinfo {volume} {22}},\ \bibinfo {pages} {315302} (\bibinfo {year} {2011})}\BibitemShut {NoStop}%
\bibitem [{\citenamefont {Pauling}(1931)}]{Pauling1931}%
  \BibitemOpen
  \bibfield  {author} {\bibinfo {author} {\bibfnamefont {L.}~\bibnamefont {Pauling}},\ }\bibfield  {title} {\bibinfo {title} {The nature of the chemical bond. application of results obtained from the quantum mechanics and from a theory of paramagnetic susceptibility to the structure of molecules},\ }\href {https://doi.org/10.1021/ja01355a027} {\bibfield  {journal} {\bibinfo  {journal} {Journal of the American Chemical Society}\ }\textbf {\bibinfo {volume} {53}},\ \bibinfo {pages} {1367–1400} (\bibinfo {year} {1931})}\BibitemShut {NoStop}%
\bibitem [{\citenamefont {Zhang}\ \emph {et~al.}(2022{\natexlab{a}})\citenamefont {Zhang}, \citenamefont {Zheng}, \citenamefont {Ma}, \citenamefont {Jiang},\ and\ \citenamefont {Li}}]{OxidationReview}%
  \BibitemOpen
  \bibfield  {author} {\bibinfo {author} {\bibfnamefont {X.}~\bibnamefont {Zhang}}, \bibinfo {author} {\bibfnamefont {P.}~\bibnamefont {Zheng}}, \bibinfo {author} {\bibfnamefont {Y.}~\bibnamefont {Ma}}, \bibinfo {author} {\bibfnamefont {Y.}~\bibnamefont {Jiang}},\ and\ \bibinfo {author} {\bibfnamefont {H.}~\bibnamefont {Li}},\ }\bibfield  {title} {\bibinfo {title} {Atomic-scale understanding of oxidation mechanisms of materials by computational approaches: A review},\ }\href@noop {} {\bibfield  {journal} {\bibinfo  {journal} {Materials\& Design}\ }\textbf {\bibinfo {volume} {217}},\ \bibinfo {pages} {110605} (\bibinfo {year} {2022}{\natexlab{a}})}\BibitemShut {NoStop}%
\bibitem [{\citenamefont {Cabrera}\ and\ \citenamefont {Mott}(1949)}]{Cabrera_1949}%
  \BibitemOpen
  \bibfield  {author} {\bibinfo {author} {\bibfnamefont {N.}~\bibnamefont {Cabrera}}\ and\ \bibinfo {author} {\bibfnamefont {N.~F.}\ \bibnamefont {Mott}},\ }\bibfield  {title} {\bibinfo {title} {Theory of the oxidation of metals},\ }\href {https://doi.org/10.1088/0034-4885/12/1/308} {\bibfield  {journal} {\bibinfo  {journal} {Reports on Progress in Physics}\ }\textbf {\bibinfo {volume} {12}},\ \bibinfo {pages} {163} (\bibinfo {year} {1949})}\BibitemShut {NoStop}%
\bibitem [{\citenamefont {Xu}\ \emph {et~al.}(2012)\citenamefont {Xu}, \citenamefont {Rosso},\ and\ \citenamefont {Bruemmer}}]{OxidationKinetics2012}%
  \BibitemOpen
  \bibfield  {author} {\bibinfo {author} {\bibfnamefont {Z.}~\bibnamefont {Xu}}, \bibinfo {author} {\bibfnamefont {K.~M.}\ \bibnamefont {Rosso}},\ and\ \bibinfo {author} {\bibfnamefont {S.}~\bibnamefont {Bruemmer}},\ }\bibfield  {title} {\bibinfo {title} {Metal oxidation kinetics and the transition from thin to thick films},\ }\href {https://doi.org/10.1039/C2CP42760E} {\bibfield  {journal} {\bibinfo  {journal} {Phys. Chem. Chem. Phys.}\ }\textbf {\bibinfo {volume} {14}},\ \bibinfo {pages} {14534} (\bibinfo {year} {2012})}\BibitemShut {NoStop}%
\bibitem [{\citenamefont {Cai}\ \emph {et~al.}(2012)\citenamefont {Cai}, \citenamefont {Zhou}, \citenamefont {Müller},\ and\ \citenamefont {Starr}}]{TemperatureAndPressureMottPotentials2012}%
  \BibitemOpen
  \bibfield  {author} {\bibinfo {author} {\bibfnamefont {N.}~\bibnamefont {Cai}}, \bibinfo {author} {\bibfnamefont {G.}~\bibnamefont {Zhou}}, \bibinfo {author} {\bibfnamefont {K.}~\bibnamefont {Müller}},\ and\ \bibinfo {author} {\bibfnamefont {D.~E.}\ \bibnamefont {Starr}},\ }\bibfield  {title} {\bibinfo {title} {{Temperature and pressure dependent Mott potentials and their influence on self-limiting oxide film growth}},\ }\href {https://doi.org/10.1063/1.4764552} {\bibfield  {journal} {\bibinfo  {journal} {Applied Physics Letters}\ }\textbf {\bibinfo {volume} {101}},\ \bibinfo {pages} {171605} (\bibinfo {year} {2012})},\ \Eprint {https://arxiv.org/abs/https://pubs.aip.org/aip/apl/article-pdf/doi/10.1063/1.4764552/14257769/171605\_1\_online.pdf} {https://pubs.aip.org/aip/apl/article-pdf/doi/10.1063/1.4764552/14257769/171605\_1\_online.pdf} \BibitemShut {NoStop}%
\bibitem [{\citenamefont {Mallison}(1995)}]{Mallison1995}%
  \BibitemOpen
  \bibfield  {author} {\bibinfo {author} {\bibfnamefont {W.~H.}\ \bibnamefont {Mallison}},\ }\bibfield  {title} {\bibinfo {title} {{Dependence of Critical Current Density on Oxygen Exposure in Nb-AlOx-Nb Tunnel Junctions}},\ }\href {https://doi.org/10.1109/77.384565} {\bibfield  {journal} {\bibinfo  {journal} {IEEE Transactions on Applied Superconductivity}\ }\textbf {\bibinfo {volume} {5}},\ \bibinfo {pages} {26} (\bibinfo {year} {1995})}\BibitemShut {NoStop}%
\bibitem [{\citenamefont {Kim}\ \emph {et~al.}(2020)\citenamefont {Kim}, \citenamefont {Ray},\ and\ \citenamefont {Lordi}}]{AlOxJunctionsLordiAPL2020}%
  \BibitemOpen
  \bibfield  {author} {\bibinfo {author} {\bibfnamefont {C.-E.}\ \bibnamefont {Kim}}, \bibinfo {author} {\bibfnamefont {K.~G.}\ \bibnamefont {Ray}},\ and\ \bibinfo {author} {\bibfnamefont {V.}~\bibnamefont {Lordi}},\ }\bibfield  {title} {\bibinfo {title} {{A density-functional theory study of the Al/AlOx/Al tunnel junction}},\ }\href {https://doi.org/10.1063/5.0020292} {\bibfield  {journal} {\bibinfo  {journal} {Journal of Applied Physics}\ }\textbf {\bibinfo {volume} {128}},\ \bibinfo {pages} {155102} (\bibinfo {year} {2020})},\ \Eprint {https://arxiv.org/abs/https://pubs.aip.org/aip/jap/article-pdf/doi/10.1063/5.0020292/14113892/155102\_1\_online.pdf} {https://pubs.aip.org/aip/jap/article-pdf/doi/10.1063/5.0020292/14113892/155102\_1\_online.pdf} \BibitemShut {NoStop}%
\bibitem [{\citenamefont {Cyster}\ \emph {et~al.}(2021)\citenamefont {Cyster}, \citenamefont {Smith}, \citenamefont {Vogt}, \citenamefont {Opletal}, \citenamefont {Russo},\ and\ \citenamefont {Cole}}]{AlOxSimulationsCysterNPJQI2021}%
  \BibitemOpen
  \bibfield  {author} {\bibinfo {author} {\bibfnamefont {M.~J.}\ \bibnamefont {Cyster}}, \bibinfo {author} {\bibfnamefont {J.~S.}\ \bibnamefont {Smith}}, \bibinfo {author} {\bibfnamefont {N.}~\bibnamefont {Vogt}}, \bibinfo {author} {\bibfnamefont {G.}~\bibnamefont {Opletal}}, \bibinfo {author} {\bibfnamefont {S.~P.}\ \bibnamefont {Russo}},\ and\ \bibinfo {author} {\bibfnamefont {J.~H.}\ \bibnamefont {Cole}},\ }\bibfield  {title} {\bibinfo {title} {Simulating the fabrication of aluminium oxide tunnel junctions},\ }\href {https://doi.org/10.1038/s41534-020-00360-4} {\bibfield  {journal} {\bibinfo  {journal} {npj Quantum Information}\ }\textbf {\bibinfo {volume} {7}},\ \bibinfo {pages} {12} (\bibinfo {year} {2021})}\BibitemShut {NoStop}%
\bibitem [{\citenamefont {Pishchimova}\ \emph {et~al.}(2023)\citenamefont {Pishchimova}, \citenamefont {Smirnov}, \citenamefont {Ezenkova}, \citenamefont {Krivko}, \citenamefont {Zikiy}, \citenamefont {Moskalev}, \citenamefont {Ivanov}, \citenamefont {Korshakov},\ and\ \citenamefont {Rodionov}}]{ImprovingJunctionReproducibilityPishchimovaSciReports2023}%
  \BibitemOpen
  \bibfield  {author} {\bibinfo {author} {\bibfnamefont {A.~A.}\ \bibnamefont {Pishchimova}}, \bibinfo {author} {\bibfnamefont {N.~S.}\ \bibnamefont {Smirnov}}, \bibinfo {author} {\bibfnamefont {D.~A.}\ \bibnamefont {Ezenkova}}, \bibinfo {author} {\bibfnamefont {E.~A.}\ \bibnamefont {Krivko}}, \bibinfo {author} {\bibfnamefont {E.~V.}\ \bibnamefont {Zikiy}}, \bibinfo {author} {\bibfnamefont {D.~O.}\ \bibnamefont {Moskalev}}, \bibinfo {author} {\bibfnamefont {A.~I.}\ \bibnamefont {Ivanov}}, \bibinfo {author} {\bibfnamefont {N.~D.}\ \bibnamefont {Korshakov}},\ and\ \bibinfo {author} {\bibfnamefont {I.~A.}\ \bibnamefont {Rodionov}},\ }\bibfield  {title} {\bibinfo {title} {Improving josephson junction reproducibility for superconducting quantum circuits: junction area fluctuation},\ }\bibfield  {journal} {\bibinfo  {journal} {Scientific Reports}\ }\textbf {\bibinfo {volume} {13}},\ \href {https://doi.org/10.1038/s41598-023-34051-9} {10.1038/s41598-023-34051-9} (\bibinfo {year} {2023})\BibitemShut {NoStop}%
\bibitem [{\citenamefont {Vettoliere}\ \emph {et~al.}(2020)\citenamefont {Vettoliere}, \citenamefont {Talamo}, \citenamefont {Silvestrini}, \citenamefont {Valentino},\ and\ \citenamefont {Granata}}]{AnnealingWafersNbAlAlOxNbVettoliere2020}%
  \BibitemOpen
  \bibfield  {author} {\bibinfo {author} {\bibfnamefont {A.}~\bibnamefont {Vettoliere}}, \bibinfo {author} {\bibfnamefont {O.}~\bibnamefont {Talamo}}, \bibinfo {author} {\bibfnamefont {P.}~\bibnamefont {Silvestrini}}, \bibinfo {author} {\bibfnamefont {M.}~\bibnamefont {Valentino}},\ and\ \bibinfo {author} {\bibfnamefont {C.}~\bibnamefont {Granata}},\ }\bibfield  {title} {\bibinfo {title} {Fine optimization of josephson critical current in squid devices by thermal annealing},\ }\href {https://doi.org/10.1088/1742-6596/1559/1/012014} {\bibfield  {journal} {\bibinfo  {journal} {Journal of Physics: Conference Series}\ }\textbf {\bibinfo {volume} {1559}},\ \bibinfo {pages} {012014} (\bibinfo {year} {2020})}\BibitemShut {NoStop}%
\bibitem [{\citenamefont {Granata}\ \emph {et~al.}(2007)\citenamefont {Granata}, \citenamefont {Vettoliere}, \citenamefont {Petti}, \citenamefont {Rippa}, \citenamefont {Ruggiero}, \citenamefont {Mormile},\ and\ \citenamefont {Russo}}]{LocalizedAnnealingNbAlAlOxNbGranata2007}%
  \BibitemOpen
  \bibfield  {author} {\bibinfo {author} {\bibfnamefont {C.}~\bibnamefont {Granata}}, \bibinfo {author} {\bibfnamefont {A.}~\bibnamefont {Vettoliere}}, \bibinfo {author} {\bibfnamefont {L.}~\bibnamefont {Petti}}, \bibinfo {author} {\bibfnamefont {M.}~\bibnamefont {Rippa}}, \bibinfo {author} {\bibfnamefont {B.}~\bibnamefont {Ruggiero}}, \bibinfo {author} {\bibfnamefont {P.}~\bibnamefont {Mormile}},\ and\ \bibinfo {author} {\bibfnamefont {M.}~\bibnamefont {Russo}},\ }\bibfield  {title} {\bibinfo {title} {Localized laser trimming of critical current in niobium based josephson devices},\ }\bibfield  {journal} {\bibinfo  {journal} {Applied Physics Letters}\ }\textbf {\bibinfo {volume} {90}},\ \href {https://doi.org/10.1063/1.2746060} {10.1063/1.2746060} (\bibinfo {year} {2007})\BibitemShut {NoStop}%
\bibitem [{\citenamefont {Hertzberg}\ \emph {et~al.}(2021)\citenamefont {Hertzberg}, \citenamefont {Zhang}, \citenamefont {Rosenblatt}, \citenamefont {Magesan}, \citenamefont {Smolin}, \citenamefont {Yau}, \citenamefont {Adiga}, \citenamefont {Sandberg}, \citenamefont {Brink}, \citenamefont {Chow},\ and\ \citenamefont {Orcutt}}]{Hertzberg2021}%
  \BibitemOpen
  \bibfield  {author} {\bibinfo {author} {\bibfnamefont {J.~B.}\ \bibnamefont {Hertzberg}}, \bibinfo {author} {\bibfnamefont {E.~J.}\ \bibnamefont {Zhang}}, \bibinfo {author} {\bibfnamefont {S.}~\bibnamefont {Rosenblatt}}, \bibinfo {author} {\bibfnamefont {E.}~\bibnamefont {Magesan}}, \bibinfo {author} {\bibfnamefont {J.~A.}\ \bibnamefont {Smolin}}, \bibinfo {author} {\bibfnamefont {J.-B.}\ \bibnamefont {Yau}}, \bibinfo {author} {\bibfnamefont {V.~P.}\ \bibnamefont {Adiga}}, \bibinfo {author} {\bibfnamefont {M.}~\bibnamefont {Sandberg}}, \bibinfo {author} {\bibfnamefont {M.}~\bibnamefont {Brink}}, \bibinfo {author} {\bibfnamefont {J.~M.}\ \bibnamefont {Chow}},\ and\ \bibinfo {author} {\bibfnamefont {J.~S.}\ \bibnamefont {Orcutt}},\ }\bibfield  {title} {\bibinfo {title} {Laser-annealing josephson junctions for yielding scaled-up superconducting quantum processors},\ }\href {https://doi.org/10.1038/s41534-021-00464-5} {\bibfield  {journal} {\bibinfo  {journal} {npj Quantum Information}\ }\textbf {\bibinfo
  {volume} {7}},\ \bibinfo {pages} {129} (\bibinfo {year} {2021})}\BibitemShut {NoStop}%
\bibitem [{\citenamefont {Zhang}\ \emph {et~al.}(2022{\natexlab{b}})\citenamefont {Zhang}, \citenamefont {Srinivasan}, \citenamefont {Sundaresan}, \citenamefont {Bogorin}, \citenamefont {Martin}, \citenamefont {Hertzberg}, \citenamefont {Timmerwilke}, \citenamefont {Pritchett}, \citenamefont {Yau}, \citenamefont {Wang}, \citenamefont {Landers}, \citenamefont {Lewandowski}, \citenamefont {Narasgond}, \citenamefont {Rosenblatt}, \citenamefont {Keefe}, \citenamefont {Lauer}, \citenamefont {Rothwell}, \citenamefont {McClure}, \citenamefont {Dial}, \citenamefont {Orcutt}, \citenamefont {Brink},\ and\ \citenamefont {Chow}}]{LasiQScience2022}%
  \BibitemOpen
  \bibfield  {author} {\bibinfo {author} {\bibfnamefont {E.~J.}\ \bibnamefont {Zhang}}, \bibinfo {author} {\bibfnamefont {S.}~\bibnamefont {Srinivasan}}, \bibinfo {author} {\bibfnamefont {N.}~\bibnamefont {Sundaresan}}, \bibinfo {author} {\bibfnamefont {D.~F.}\ \bibnamefont {Bogorin}}, \bibinfo {author} {\bibfnamefont {Y.}~\bibnamefont {Martin}}, \bibinfo {author} {\bibfnamefont {J.~B.}\ \bibnamefont {Hertzberg}}, \bibinfo {author} {\bibfnamefont {J.}~\bibnamefont {Timmerwilke}}, \bibinfo {author} {\bibfnamefont {E.~J.}\ \bibnamefont {Pritchett}}, \bibinfo {author} {\bibfnamefont {J.-B.}\ \bibnamefont {Yau}}, \bibinfo {author} {\bibfnamefont {C.}~\bibnamefont {Wang}}, \bibinfo {author} {\bibfnamefont {W.}~\bibnamefont {Landers}}, \bibinfo {author} {\bibfnamefont {E.~P.}\ \bibnamefont {Lewandowski}}, \bibinfo {author} {\bibfnamefont {A.}~\bibnamefont {Narasgond}}, \bibinfo {author} {\bibfnamefont {S.}~\bibnamefont {Rosenblatt}}, \bibinfo {author} {\bibfnamefont {G.~A.}\ \bibnamefont {Keefe}}, \bibinfo {author}
  {\bibfnamefont {I.}~\bibnamefont {Lauer}}, \bibinfo {author} {\bibfnamefont {M.~B.}\ \bibnamefont {Rothwell}}, \bibinfo {author} {\bibfnamefont {D.~T.}\ \bibnamefont {McClure}}, \bibinfo {author} {\bibfnamefont {O.~E.}\ \bibnamefont {Dial}}, \bibinfo {author} {\bibfnamefont {J.~S.}\ \bibnamefont {Orcutt}}, \bibinfo {author} {\bibfnamefont {M.}~\bibnamefont {Brink}},\ and\ \bibinfo {author} {\bibfnamefont {J.~M.}\ \bibnamefont {Chow}},\ }\bibfield  {title} {\bibinfo {title} {High-performance superconducting quantum processors via laser annealing of transmon qubits},\ }\href {https://doi.org/10.1126/sciadv.abi6690} {\bibfield  {journal} {\bibinfo  {journal} {Science Advances}\ }\textbf {\bibinfo {volume} {8}},\ \bibinfo {pages} {eabi6690} (\bibinfo {year} {2022}{\natexlab{b}})},\ \Eprint {https://arxiv.org/abs/https://www.science.org/doi/pdf/10.1126/sciadv.abi6690} {https://www.science.org/doi/pdf/10.1126/sciadv.abi6690} \BibitemShut {NoStop}%
\bibitem [{\citenamefont {Kim}\ \emph {et~al.}(2022)\citenamefont {Kim}, \citenamefont {Jünger}, \citenamefont {Morvan}, \citenamefont {Barnard}, \citenamefont {Livingston}, \citenamefont {Altoé}, \citenamefont {Kim}, \citenamefont {Song}, \citenamefont {Chen}, \citenamefont {Kreikebaum}, \citenamefont {Ogletree}, \citenamefont {Santiago},\ and\ \citenamefont {Siddiqi}}]{LasiQBerkeley2022}%
  \BibitemOpen
  \bibfield  {author} {\bibinfo {author} {\bibfnamefont {H.}~\bibnamefont {Kim}}, \bibinfo {author} {\bibfnamefont {C.}~\bibnamefont {Jünger}}, \bibinfo {author} {\bibfnamefont {A.}~\bibnamefont {Morvan}}, \bibinfo {author} {\bibfnamefont {E.~S.}\ \bibnamefont {Barnard}}, \bibinfo {author} {\bibfnamefont {W.~P.}\ \bibnamefont {Livingston}}, \bibinfo {author} {\bibfnamefont {M.~V.~P.}\ \bibnamefont {Altoé}}, \bibinfo {author} {\bibfnamefont {Y.}~\bibnamefont {Kim}}, \bibinfo {author} {\bibfnamefont {C.}~\bibnamefont {Song}}, \bibinfo {author} {\bibfnamefont {L.}~\bibnamefont {Chen}}, \bibinfo {author} {\bibfnamefont {J.~M.}\ \bibnamefont {Kreikebaum}}, \bibinfo {author} {\bibfnamefont {D.~F.}\ \bibnamefont {Ogletree}}, \bibinfo {author} {\bibfnamefont {D.~I.}\ \bibnamefont {Santiago}},\ and\ \bibinfo {author} {\bibfnamefont {I.}~\bibnamefont {Siddiqi}},\ }\bibfield  {title} {\bibinfo {title} {{Effects of laser-annealing on fixed-frequency superconducting qubits}},\ }\href {https://doi.org/10.1063/5.0102092}
  {\bibfield  {journal} {\bibinfo  {journal} {Applied Physics Letters}\ }\textbf {\bibinfo {volume} {121}},\ \bibinfo {pages} {142601} (\bibinfo {year} {2022})},\ \Eprint {https://arxiv.org/abs/https://pubs.aip.org/aip/apl/article-pdf/doi/10.1063/5.0102092/16484380/142601\_1\_online.pdf} {https://pubs.aip.org/aip/apl/article-pdf/doi/10.1063/5.0102092/16484380/142601\_1\_online.pdf} \BibitemShut {NoStop}%
\bibitem [{\citenamefont {Balaji}\ \emph {et~al.}(2024)\citenamefont {Balaji}, \citenamefont {Acharya}, \citenamefont {Armstrong}, \citenamefont {Crawford}, \citenamefont {Danilin}, \citenamefont {Dixon}, \citenamefont {Kennedy}, \citenamefont {Pothuraju}, \citenamefont {Shahbazi},\ and\ \citenamefont {Shelly}}]{ebeamAnnealing}%
  \BibitemOpen
  \bibfield  {author} {\bibinfo {author} {\bibfnamefont {Y.}~\bibnamefont {Balaji}}, \bibinfo {author} {\bibfnamefont {N.}~\bibnamefont {Acharya}}, \bibinfo {author} {\bibfnamefont {R.}~\bibnamefont {Armstrong}}, \bibinfo {author} {\bibfnamefont {K.~G.}\ \bibnamefont {Crawford}}, \bibinfo {author} {\bibfnamefont {S.}~\bibnamefont {Danilin}}, \bibinfo {author} {\bibfnamefont {T.}~\bibnamefont {Dixon}}, \bibinfo {author} {\bibfnamefont {O.~W.}\ \bibnamefont {Kennedy}}, \bibinfo {author} {\bibfnamefont {R.~D.}\ \bibnamefont {Pothuraju}}, \bibinfo {author} {\bibfnamefont {K.}~\bibnamefont {Shahbazi}},\ and\ \bibinfo {author} {\bibfnamefont {C.~D.}\ \bibnamefont {Shelly}},\ }\href@noop {} {\bibinfo {title} {Electron-beam annealing of josephson junctions for frequency tuning of quantum processors}} (\bibinfo {year} {2024}),\ \Eprint {https://arxiv.org/abs/2402.17395} {arXiv:2402.17395 [quant-ph]} \BibitemShut {NoStop}%
\bibitem [{\citenamefont {Vallés-Sanclemente}\ \emph {et~al.}(2023)\citenamefont {Vallés-Sanclemente}, \citenamefont {van~der Meer}, \citenamefont {Finkel}, \citenamefont {Muthusubramanian}, \citenamefont {Beekman}, \citenamefont {Ali}, \citenamefont {Marques}, \citenamefont {Zachariadis}, \citenamefont {Veen}, \citenamefont {Stavenga}, \citenamefont {Haider},\ and\ \citenamefont {DiCarlo}}]{LaserAnnealDelft_1}%
  \BibitemOpen
  \bibfield  {author} {\bibinfo {author} {\bibfnamefont {S.}~\bibnamefont {Vallés-Sanclemente}}, \bibinfo {author} {\bibfnamefont {S.~L.~M.}\ \bibnamefont {van~der Meer}}, \bibinfo {author} {\bibfnamefont {M.}~\bibnamefont {Finkel}}, \bibinfo {author} {\bibfnamefont {N.}~\bibnamefont {Muthusubramanian}}, \bibinfo {author} {\bibfnamefont {M.}~\bibnamefont {Beekman}}, \bibinfo {author} {\bibfnamefont {H.}~\bibnamefont {Ali}}, \bibinfo {author} {\bibfnamefont {J.~F.}\ \bibnamefont {Marques}}, \bibinfo {author} {\bibfnamefont {C.}~\bibnamefont {Zachariadis}}, \bibinfo {author} {\bibfnamefont {H.~M.}\ \bibnamefont {Veen}}, \bibinfo {author} {\bibfnamefont {T.}~\bibnamefont {Stavenga}}, \bibinfo {author} {\bibfnamefont {N.}~\bibnamefont {Haider}},\ and\ \bibinfo {author} {\bibfnamefont {L.}~\bibnamefont {DiCarlo}},\ }\bibfield  {title} {\bibinfo {title} {{Post-fabrication frequency trimming of coplanar-waveguide resonators in circuit QED quantum processors}},\ }\href {https://doi.org/10.1063/5.0148222} {\bibfield
  {journal} {\bibinfo  {journal} {Applied Physics Letters}\ }\textbf {\bibinfo {volume} {123}},\ \bibinfo {pages} {034004} (\bibinfo {year} {2023})},\ \Eprint {https://arxiv.org/abs/https://pubs.aip.org/aip/apl/article-pdf/doi/10.1063/5.0148222/18051172/034004\_1\_5.0148222.pdf} {https://pubs.aip.org/aip/apl/article-pdf/doi/10.1063/5.0148222/18051172/034004\_1\_5.0148222.pdf} \BibitemShut {NoStop}%
\bibitem [{\citenamefont {Ebinger}\ and\ \citenamefont {Yates}(1998)}]{EbingerOxidationPRB1976}%
  \BibitemOpen
  \bibfield  {author} {\bibinfo {author} {\bibfnamefont {H.~D.}\ \bibnamefont {Ebinger}}\ and\ \bibinfo {author} {\bibfnamefont {J.~T.}\ \bibnamefont {Yates}},\ }\bibfield  {title} {\bibinfo {title} {Electron-impact-induced oxidation of \uppercase{A}l(111) in water vapor: Relation to the \uppercase{C}abrera-\uppercase{M}ott mechanism},\ }\href {https://doi.org/10.1103/PhysRevB.57.1976} {\bibfield  {journal} {\bibinfo  {journal} {Phys. Rev. B}\ }\textbf {\bibinfo {volume} {57}},\ \bibinfo {pages} {1976} (\bibinfo {year} {1998})}\BibitemShut {NoStop}%
\bibitem [{\citenamefont {Konkin}\ and\ \citenamefont {Adler}(1980)}]{voltageAnnealing}%
  \BibitemOpen
  \bibfield  {author} {\bibinfo {author} {\bibfnamefont {M.~K.}\ \bibnamefont {Konkin}}\ and\ \bibinfo {author} {\bibfnamefont {J.~G.}\ \bibnamefont {Adler}},\ }\bibfield  {title} {\bibinfo {title} {{Annealing effects in tunnel junctions (voltage annealing)}},\ }\href {https://doi.org/10.1063/1.327501} {\bibfield  {journal} {\bibinfo  {journal} {Journal of Applied Physics}\ }\textbf {\bibinfo {volume} {51}},\ \bibinfo {pages} {5450} (\bibinfo {year} {1980})},\ \Eprint {https://arxiv.org/abs/https://pubs.aip.org/aip/jap/article-pdf/51/10/5450/7969138/5450\_1\_online.pdf} {https://pubs.aip.org/aip/jap/article-pdf/51/10/5450/7969138/5450\_1\_online.pdf} \BibitemShut {NoStop}%
\bibitem [{\citenamefont {Zhang}(2001)}]{MagneticAnnealing}%
  \BibitemOpen
  \bibfield  {author} {\bibinfo {author} {\bibfnamefont {Y.}~\bibnamefont {Zhang}},\ }\href {https://patents.google.com/patent/US6217672B1/en?oq=us6217672+b1} {\bibinfo {title} {Magnetic annealing of magnetic alloys in a dynamic magnetic field}} (\bibinfo {year} {2001}),\ \bibinfo {note} {uS Patent 6,217,672 B1}\BibitemShut {NoStop}%
\bibitem [{\citenamefont {Yamamoto}\ \emph {et~al.}(2014)\citenamefont {Yamamoto}, \citenamefont {Yamashita},\ and\ \citenamefont {et. al}}]{PiezoElectricAnnealing}%
  \BibitemOpen
  \bibfield  {author} {\bibinfo {author} {\bibfnamefont {N.}~\bibnamefont {Yamamoto}}, \bibinfo {author} {\bibnamefont {Yamashita}},\ and\ \bibinfo {author} {\bibnamefont {et. al}},\ }\href {https://patents.google.com/patent/US20140062261A1/en?oq=us2014%2f0062261} {\bibinfo {title} {Ultrasonic probe, piezoelectric transducer, method of manufacturing ultrasonic probe, and method of manufacturing piezoelectric transducer}} (\bibinfo {year} {2014}),\ \bibinfo {note} {uS Patent 2014/0062261 A1}\BibitemShut {NoStop}%
\bibitem [{\citenamefont {Fontanini}\ \emph {et~al.}(2022)\citenamefont {Fontanini}, \citenamefont {Barbot}, \citenamefont {Segatto}, \citenamefont {Lancaster}, \citenamefont {Duong}, \citenamefont {Driussi}, \citenamefont {Grenouillet}, \citenamefont {Triozon}, \citenamefont {Coignus}, \citenamefont {Mikolajick}, \citenamefont {Slesazeck},\ and\ \citenamefont {Esseni}}]{ftj}%
  \BibitemOpen
  \bibfield  {author} {\bibinfo {author} {\bibfnamefont {R.}~\bibnamefont {Fontanini}}, \bibinfo {author} {\bibfnamefont {J.}~\bibnamefont {Barbot}}, \bibinfo {author} {\bibfnamefont {M.}~\bibnamefont {Segatto}}, \bibinfo {author} {\bibfnamefont {S.}~\bibnamefont {Lancaster}}, \bibinfo {author} {\bibfnamefont {Q.}~\bibnamefont {Duong}}, \bibinfo {author} {\bibfnamefont {F.}~\bibnamefont {Driussi}}, \bibinfo {author} {\bibfnamefont {L.}~\bibnamefont {Grenouillet}}, \bibinfo {author} {\bibfnamefont {L.}~\bibnamefont {Triozon}}, \bibinfo {author} {\bibfnamefont {J.}~\bibnamefont {Coignus}}, \bibinfo {author} {\bibfnamefont {T.}~\bibnamefont {Mikolajick}}, \bibinfo {author} {\bibfnamefont {S.}~\bibnamefont {Slesazeck}},\ and\ \bibinfo {author} {\bibfnamefont {D.}~\bibnamefont {Esseni}},\ }\bibfield  {title} {\bibinfo {title} {Interplay between charge trapping and polarization switching in beol-compatible bilayer ferroelectric tunnel junctions},\ }\href {https://doi.org/10.1109/JEDS.2022.3171217} {\bibfield
  {journal} {\bibinfo  {journal} {IEEE Journal of the Electron Devices Society}\ }\textbf {\bibinfo {volume} {10}},\ \bibinfo {pages} {1} (\bibinfo {year} {2022})}\BibitemShut {NoStop}%
\bibitem [{\citenamefont {Konkin}\ and\ \citenamefont {Adler}(1982)}]{APAA_Konkin_1982}%
  \BibitemOpen
  \bibfield  {author} {\bibinfo {author} {\bibfnamefont {M.~K.}\ \bibnamefont {Konkin}}\ and\ \bibinfo {author} {\bibfnamefont {J.~G.}\ \bibnamefont {Adler}},\ }\bibfield  {title} {\bibinfo {title} {{Annealing effects in tunnel junctions (voltage annealing with alternating polarity)}},\ }\href {https://doi.org/10.1063/1.331337} {\bibfield  {journal} {\bibinfo  {journal} {Journal of Applied Physics}\ }\textbf {\bibinfo {volume} {53}},\ \bibinfo {pages} {5057} (\bibinfo {year} {1982})},\ \Eprint {https://arxiv.org/abs/https://pubs.aip.org/aip/jap/article-pdf/53/7/5057/18397790/5057\_1\_online.pdf} {https://pubs.aip.org/aip/jap/article-pdf/53/7/5057/18397790/5057\_1\_online.pdf} \BibitemShut {NoStop}%
\bibitem [{\citenamefont {Žurauskienė}\ \emph {et~al.}(2013)\citenamefont {Žurauskienė}, \citenamefont {Balevičius}, \citenamefont {Pavilonis}, \citenamefont {Stankevič}, \citenamefont {Keršulis},\ and\ \citenamefont {Novickij}}]{MagneticRelaxation}%
  \BibitemOpen
  \bibfield  {author} {\bibinfo {author} {\bibfnamefont {N.}~\bibnamefont {Žurauskienė}}, \bibinfo {author} {\bibfnamefont {S.}~\bibnamefont {Balevičius}}, \bibinfo {author} {\bibfnamefont {D.}~\bibnamefont {Pavilonis}}, \bibinfo {author} {\bibfnamefont {V.}~\bibnamefont {Stankevič}}, \bibinfo {author} {\bibfnamefont {S.}~\bibnamefont {Keršulis}},\ and\ \bibinfo {author} {\bibfnamefont {J.}~\bibnamefont {Novickij}},\ }\bibfield  {title} {\bibinfo {title} {Magnetoresistance relaxation in thin la-sr-mn-o films exposed to high-pulsed magnetic fields},\ }\href {https://doi.org/10.1109/TPS.2013.2261558} {\bibfield  {journal} {\bibinfo  {journal} {IEEE Transactions on Plasma Science}\ }\textbf {\bibinfo {volume} {41}},\ \bibinfo {pages} {2830} (\bibinfo {year} {2013})}\BibitemShut {NoStop}%
\bibitem [{\citenamefont {Hashimoto}\ \emph {et~al.}(2022)\citenamefont {Hashimoto}, \citenamefont {Onodera}, \citenamefont {Tahara}, \citenamefont {Kohara}, \citenamefont {Yazawa}, \citenamefont {Segawa}, \citenamefont {Murakami},\ and\ \citenamefont {Ohara}}]{Al2O3GlassTransitionTHashimoto2022}%
  \BibitemOpen
  \bibfield  {author} {\bibinfo {author} {\bibfnamefont {H.}~\bibnamefont {Hashimoto}}, \bibinfo {author} {\bibfnamefont {Y.}~\bibnamefont {Onodera}}, \bibinfo {author} {\bibfnamefont {S.}~\bibnamefont {Tahara}}, \bibinfo {author} {\bibfnamefont {S.}~\bibnamefont {Kohara}}, \bibinfo {author} {\bibfnamefont {K.}~\bibnamefont {Yazawa}}, \bibinfo {author} {\bibfnamefont {H.}~\bibnamefont {Segawa}}, \bibinfo {author} {\bibfnamefont {M.}~\bibnamefont {Murakami}},\ and\ \bibinfo {author} {\bibfnamefont {K.}~\bibnamefont {Ohara}},\ }\bibfield  {title} {\bibinfo {title} {Structure of alumina glass},\ }\href {https://doi.org/10.1038/s41598-021-04455-6} {\bibfield  {journal} {\bibinfo  {journal} {Scientific Reports}\ }\textbf {\bibinfo {volume} {12}},\ \bibinfo {pages} {516} (\bibinfo {year} {2022})}\BibitemShut {NoStop}%
\bibitem [{\citenamefont {Manenti}\ \emph {et~al.}(2021)\citenamefont {Manenti}, \citenamefont {Sete}, \citenamefont {Chen}, \citenamefont {Kulshreshtha}, \citenamefont {Yeh}, \citenamefont {Oruc}, \citenamefont {Bestwick}, \citenamefont {Field}, \citenamefont {Jackson},\ and\ \citenamefont {Poletto}}]{RigettiQubits}%
  \BibitemOpen
  \bibfield  {author} {\bibinfo {author} {\bibfnamefont {R.}~\bibnamefont {Manenti}}, \bibinfo {author} {\bibfnamefont {E.~A.}\ \bibnamefont {Sete}}, \bibinfo {author} {\bibfnamefont {A.~Q.}\ \bibnamefont {Chen}}, \bibinfo {author} {\bibfnamefont {S.}~\bibnamefont {Kulshreshtha}}, \bibinfo {author} {\bibfnamefont {J.-H.}\ \bibnamefont {Yeh}}, \bibinfo {author} {\bibfnamefont {F.}~\bibnamefont {Oruc}}, \bibinfo {author} {\bibfnamefont {A.}~\bibnamefont {Bestwick}}, \bibinfo {author} {\bibfnamefont {M.}~\bibnamefont {Field}}, \bibinfo {author} {\bibfnamefont {K.}~\bibnamefont {Jackson}},\ and\ \bibinfo {author} {\bibfnamefont {S.}~\bibnamefont {Poletto}},\ }\bibfield  {title} {\bibinfo {title} {{Full control of superconducting qubits with combined on-chip microwave and flux lines}},\ }\href {https://doi.org/10.1063/5.0065517} {\bibfield  {journal} {\bibinfo  {journal} {Applied Physics Letters}\ }\textbf {\bibinfo {volume} {119}},\ \bibinfo {pages} {144001} (\bibinfo {year} {2021})},\ \Eprint
  {https://arxiv.org/abs/https://pubs.aip.org/aip/apl/article-pdf/doi/10.1063/5.0065517/13194475/144001\_1\_online.pdf} {https://pubs.aip.org/aip/apl/article-pdf/doi/10.1063/5.0065517/13194475/144001\_1\_online.pdf} \BibitemShut {NoStop}%
\bibitem [{\citenamefont {Bilmes}\ \emph {et~al.}(2022)\citenamefont {Bilmes}, \citenamefont {Volosheniuk}, \citenamefont {Ustinov},\ and\ \citenamefont {Lisenfeld}}]{JunctionVsSurfaceTLS_Lisenfeld2022}%
  \BibitemOpen
  \bibfield  {author} {\bibinfo {author} {\bibfnamefont {A.}~\bibnamefont {Bilmes}}, \bibinfo {author} {\bibfnamefont {S.}~\bibnamefont {Volosheniuk}}, \bibinfo {author} {\bibfnamefont {A.~V.}\ \bibnamefont {Ustinov}},\ and\ \bibinfo {author} {\bibfnamefont {J.}~\bibnamefont {Lisenfeld}},\ }\bibfield  {title} {\bibinfo {title} {Probing defect densities at the edges and inside josephson junctions of superconducting qubits},\ }\bibfield  {journal} {\bibinfo  {journal} {npj Quantum Information}\ }\textbf {\bibinfo {volume} {8}},\ \href {https://doi.org/10.1038/s41534-022-00532-4} {10.1038/s41534-022-00532-4} (\bibinfo {year} {2022})\BibitemShut {NoStop}%
\bibitem [{\citenamefont {Lisenfeld}\ \emph {et~al.}(2019)\citenamefont {Lisenfeld}, \citenamefont {Bilmes}, \citenamefont {Megrant}, \citenamefont {Barends}, \citenamefont {Kelly}, \citenamefont {Klimov}, \citenamefont {Weiss}, \citenamefont {Martinis},\ and\ \citenamefont {Ustinov}}]{JunctionVsSurfaceTLS_Lisenfeld2019}%
  \BibitemOpen
  \bibfield  {author} {\bibinfo {author} {\bibfnamefont {J.}~\bibnamefont {Lisenfeld}}, \bibinfo {author} {\bibfnamefont {A.}~\bibnamefont {Bilmes}}, \bibinfo {author} {\bibfnamefont {A.}~\bibnamefont {Megrant}}, \bibinfo {author} {\bibfnamefont {R.}~\bibnamefont {Barends}}, \bibinfo {author} {\bibfnamefont {J.}~\bibnamefont {Kelly}}, \bibinfo {author} {\bibfnamefont {P.}~\bibnamefont {Klimov}}, \bibinfo {author} {\bibfnamefont {G.}~\bibnamefont {Weiss}}, \bibinfo {author} {\bibfnamefont {J.~M.}\ \bibnamefont {Martinis}},\ and\ \bibinfo {author} {\bibfnamefont {A.~V.}\ \bibnamefont {Ustinov}},\ }\bibfield  {title} {\bibinfo {title} {Electric field spectroscopy of material defects in transmon qubits},\ }\bibfield  {journal} {\bibinfo  {journal} {npj Quantum Information}\ }\textbf {\bibinfo {volume} {5}},\ \href {https://doi.org/10.1038/s41534-019-0224-1} {10.1038/s41534-019-0224-1} (\bibinfo {year} {2019})\BibitemShut {NoStop}%
\bibitem [{\citenamefont {Nersisyan}\ \emph {et~al.}(2019)\citenamefont {Nersisyan}, \citenamefont {Poletto}, \citenamefont {Alidoust}, \citenamefont {Manenti}, \citenamefont {Renzas}, \citenamefont {Bui}, \citenamefont {Vu}, \citenamefont {Whyland}, \citenamefont {Mohan}, \citenamefont {Sete}, \citenamefont {Stanwyck}, \citenamefont {Bestwick},\ and\ \citenamefont {Reagor}}]{nersisyan2019manufacturing}%
  \BibitemOpen
  \bibfield  {author} {\bibinfo {author} {\bibfnamefont {A.}~\bibnamefont {Nersisyan}}, \bibinfo {author} {\bibfnamefont {S.}~\bibnamefont {Poletto}}, \bibinfo {author} {\bibfnamefont {N.}~\bibnamefont {Alidoust}}, \bibinfo {author} {\bibfnamefont {R.}~\bibnamefont {Manenti}}, \bibinfo {author} {\bibfnamefont {R.}~\bibnamefont {Renzas}}, \bibinfo {author} {\bibfnamefont {C.-V.}\ \bibnamefont {Bui}}, \bibinfo {author} {\bibfnamefont {K.}~\bibnamefont {Vu}}, \bibinfo {author} {\bibfnamefont {T.}~\bibnamefont {Whyland}}, \bibinfo {author} {\bibfnamefont {Y.}~\bibnamefont {Mohan}}, \bibinfo {author} {\bibfnamefont {E.~A.}\ \bibnamefont {Sete}}, \bibinfo {author} {\bibfnamefont {S.}~\bibnamefont {Stanwyck}}, \bibinfo {author} {\bibfnamefont {A.}~\bibnamefont {Bestwick}},\ and\ \bibinfo {author} {\bibfnamefont {M.}~\bibnamefont {Reagor}},\ }\href@noop {} {\bibinfo {title} {Manufacturing low dissipation superconducting quantum processors}} (\bibinfo {year} {2019}),\ \Eprint {https://arxiv.org/abs/1901.08042}
  {arXiv:1901.08042 [quant-ph]} \BibitemShut {NoStop}%
\bibitem [{\citenamefont {Wang}\ \emph {et~al.}(2015)\citenamefont {Wang}, \citenamefont {Axline}, \citenamefont {Gao}, \citenamefont {Brecht}, \citenamefont {Chu}, \citenamefont {Frunzio}, \citenamefont {Devoret},\ and\ \citenamefont {Schoelkopf}}]{wang2015surface}%
  \BibitemOpen
  \bibfield  {author} {\bibinfo {author} {\bibfnamefont {C.}~\bibnamefont {Wang}}, \bibinfo {author} {\bibfnamefont {C.}~\bibnamefont {Axline}}, \bibinfo {author} {\bibfnamefont {Y.}~\bibnamefont {Gao}}, \bibinfo {author} {\bibfnamefont {T.}~\bibnamefont {Brecht}}, \bibinfo {author} {\bibfnamefont {Y.}~\bibnamefont {Chu}}, \bibinfo {author} {\bibfnamefont {L.}~\bibnamefont {Frunzio}}, \bibinfo {author} {\bibfnamefont {M.}~\bibnamefont {Devoret}},\ and\ \bibinfo {author} {\bibfnamefont {R.}~\bibnamefont {Schoelkopf}},\ }\bibfield  {title} {\bibinfo {title} {Surface participation and dielectric loss in superconducting qubits},\ }\href@noop {} {\bibfield  {journal} {\bibinfo  {journal} {Appl. Phys. Lett.}\ }\textbf {\bibinfo {volume} {107}},\ \bibinfo {pages} {162601} (\bibinfo {year} {2015})}\BibitemShut {NoStop}%
\bibitem [{\citenamefont {Oh}\ \emph {et~al.}(2024)\citenamefont {Oh}, \citenamefont {Kopas}, \citenamefont {Canizoglu}, \citenamefont {Lachman}, \citenamefont {Yadavalli}, \citenamefont {Mutus}, \citenamefont {Kim}, \citenamefont {Kramer}, \citenamefont {King},\ and\ \citenamefont {Zhou}}]{ALD_Oh_2024}%
  \BibitemOpen
  \bibfield  {author} {\bibinfo {author} {\bibfnamefont {J.-S.}\ \bibnamefont {Oh}}, \bibinfo {author} {\bibfnamefont {C.}~\bibnamefont {Kopas}}, \bibinfo {author} {\bibfnamefont {H.}~\bibnamefont {Canizoglu}}, \bibinfo {author} {\bibfnamefont {E.}~\bibnamefont {Lachman}}, \bibinfo {author} {\bibfnamefont {K.}~\bibnamefont {Yadavalli}}, \bibinfo {author} {\bibfnamefont {J.}~\bibnamefont {Mutus}}, \bibinfo {author} {\bibfnamefont {T.-H.}\ \bibnamefont {Kim}}, \bibinfo {author} {\bibfnamefont {M.}~\bibnamefont {Kramer}}, \bibinfo {author} {\bibfnamefont {A.}~\bibnamefont {King}},\ and\ \bibinfo {author} {\bibfnamefont {L.}~\bibnamefont {Zhou}},\ }\href@noop {} {\bibinfo {title} {Correlating aluminum layer deposition rates, josephson junction microstructure, and superconducting qubits’ performance}} (\bibinfo {year} {2024}),\ \bibinfo {note} {submitted to Acta Materialia}\BibitemShut {NoStop}%
\bibitem [{\citenamefont {Bouchet}\ and\ \citenamefont {Colliex}(2003)}]{BouchetAluminaUltraMicroscopy2003}%
  \BibitemOpen
  \bibfield  {author} {\bibinfo {author} {\bibfnamefont {D.}~\bibnamefont {Bouchet}}\ and\ \bibinfo {author} {\bibfnamefont {C.}~\bibnamefont {Colliex}},\ }\bibfield  {title} {\bibinfo {title} {Experimental study of elnes at grain boundaries in alumina: intergranular radiation damage effects on al-l23 and o-k edges},\ }\href {https://doi.org/10.1016/s0304-3991(02)00437-0} {\bibfield  {journal} {\bibinfo  {journal} {Ultramicroscopy}\ }\textbf {\bibinfo {volume} {96}},\ \bibinfo {pages} {139–152} (\bibinfo {year} {2003})}\BibitemShut {NoStop}%
\bibitem [{\citenamefont {Fritz}\ \emph {et~al.}(2018)\citenamefont {Fritz}, \citenamefont {Seiler}, \citenamefont {Radtke}, \citenamefont {Schneider}, \citenamefont {Weides}, \citenamefont {Weiß},\ and\ \citenamefont {Gerthsen}}]{FritzAlOxNanostructureScientificReports2018}%
  \BibitemOpen
  \bibfield  {author} {\bibinfo {author} {\bibfnamefont {S.}~\bibnamefont {Fritz}}, \bibinfo {author} {\bibfnamefont {A.}~\bibnamefont {Seiler}}, \bibinfo {author} {\bibfnamefont {L.}~\bibnamefont {Radtke}}, \bibinfo {author} {\bibfnamefont {R.}~\bibnamefont {Schneider}}, \bibinfo {author} {\bibfnamefont {M.}~\bibnamefont {Weides}}, \bibinfo {author} {\bibfnamefont {G.}~\bibnamefont {Weiß}},\ and\ \bibinfo {author} {\bibfnamefont {D.}~\bibnamefont {Gerthsen}},\ }\bibfield  {title} {\bibinfo {title} {Correlating the nanostructure of al-oxide with deposition conditions and dielectric contributions of two-level systems in perspective of superconducting quantum circuits},\ }\bibfield  {journal} {\bibinfo  {journal} {Scientific Reports}\ }\textbf {\bibinfo {volume} {8}},\ \href {https://doi.org/10.1038/s41598-018-26066-4} {10.1038/s41598-018-26066-4} (\bibinfo {year} {2018})\BibitemShut {NoStop}%
\bibitem [{\citenamefont {Fritz}\ \emph {et~al.}(2019)\citenamefont {Fritz}, \citenamefont {Radtke}, \citenamefont {Schneider}, \citenamefont {Luysberg}, \citenamefont {Weides},\ and\ \citenamefont {Gerthsen}}]{FritzStructureMechancialAlOxPhysRevMater2019}%
  \BibitemOpen
  \bibfield  {author} {\bibinfo {author} {\bibfnamefont {S.}~\bibnamefont {Fritz}}, \bibinfo {author} {\bibfnamefont {L.}~\bibnamefont {Radtke}}, \bibinfo {author} {\bibfnamefont {R.}~\bibnamefont {Schneider}}, \bibinfo {author} {\bibfnamefont {M.}~\bibnamefont {Luysberg}}, \bibinfo {author} {\bibfnamefont {M.}~\bibnamefont {Weides}},\ and\ \bibinfo {author} {\bibfnamefont {D.}~\bibnamefont {Gerthsen}},\ }\bibfield  {title} {\bibinfo {title} {Structural and nanochemical properties of alox layers in al/alox/al-layer systems for josephson junctions},\ }\bibfield  {journal} {\bibinfo  {journal} {Physical Review Materials}\ }\textbf {\bibinfo {volume} {3}},\ \href {https://doi.org/10.1103/physrevmaterials.3.114805} {10.1103/physrevmaterials.3.114805} (\bibinfo {year} {2019})\BibitemShut {NoStop}%
\bibitem [{\citenamefont {Kimoto}\ \emph {et~al.}(2003)\citenamefont {Kimoto}, \citenamefont {Matsui}, \citenamefont {Nabatame}, \citenamefont {Yasuda}, \citenamefont {Mizoguchi}, \citenamefont {Tanaka},\ and\ \citenamefont {Toriumi}}]{KimotoCoordinationAl203APL2003}%
  \BibitemOpen
  \bibfield  {author} {\bibinfo {author} {\bibfnamefont {K.}~\bibnamefont {Kimoto}}, \bibinfo {author} {\bibfnamefont {Y.}~\bibnamefont {Matsui}}, \bibinfo {author} {\bibfnamefont {T.}~\bibnamefont {Nabatame}}, \bibinfo {author} {\bibfnamefont {T.}~\bibnamefont {Yasuda}}, \bibinfo {author} {\bibfnamefont {T.}~\bibnamefont {Mizoguchi}}, \bibinfo {author} {\bibfnamefont {I.}~\bibnamefont {Tanaka}},\ and\ \bibinfo {author} {\bibfnamefont {A.}~\bibnamefont {Toriumi}},\ }\bibfield  {title} {\bibinfo {title} {{Coordination and interface analysis of atomic-layer-deposition Al2O3 on Si(001) using energy-loss near-edge structures}},\ }\href {https://doi.org/10.1063/1.1629397} {\bibfield  {journal} {\bibinfo  {journal} {Applied Physics Letters}\ }\textbf {\bibinfo {volume} {83}},\ \bibinfo {pages} {4306} (\bibinfo {year} {2003})},\ \Eprint {https://arxiv.org/abs/https://pubs.aip.org/aip/apl/article-pdf/83/21/4306/12229282/4306\_1\_online.pdf}
  {https://pubs.aip.org/aip/apl/article-pdf/83/21/4306/12229282/4306\_1\_online.pdf} \BibitemShut {NoStop}%
\bibitem [{\citenamefont {Weigel}\ \emph {et~al.}(2008)\citenamefont {Weigel}, \citenamefont {Calas}, \citenamefont {Cormier}, \citenamefont {Galoisy},\ and\ \citenamefont {Henderson}}]{WeigelHRTEMJofPhysCondMat2008}%
  \BibitemOpen
  \bibfield  {author} {\bibinfo {author} {\bibfnamefont {C.}~\bibnamefont {Weigel}}, \bibinfo {author} {\bibfnamefont {G.}~\bibnamefont {Calas}}, \bibinfo {author} {\bibfnamefont {L.}~\bibnamefont {Cormier}}, \bibinfo {author} {\bibfnamefont {L.}~\bibnamefont {Galoisy}},\ and\ \bibinfo {author} {\bibfnamefont {G.~S.}\ \bibnamefont {Henderson}},\ }\bibfield  {title} {\bibinfo {title} {High-resolution al l2, 3-edge x-ray absorption near edge structure spectra of al-containing crystals and glasses: coordination number and bonding information from edge components},\ }\href {https://doi.org/10.1088/0953-8984/20/13/135219} {\bibfield  {journal} {\bibinfo  {journal} {Journal of Physics: Condensed Matter}\ }\textbf {\bibinfo {volume} {20}},\ \bibinfo {pages} {135219} (\bibinfo {year} {2008})}\BibitemShut {NoStop}%
\bibitem [{\citenamefont {Li}\ \emph {et~al.}(2023)\citenamefont {Li}, \citenamefont {Shen}, \citenamefont {Luo}, \citenamefont {An}, \citenamefont {Gao}, \citenamefont {Xiao},\ and\ \citenamefont {Zou}}]{DislocationMotionLiNatureMaterials2023}%
  \BibitemOpen
  \bibfield  {author} {\bibinfo {author} {\bibfnamefont {M.}~\bibnamefont {Li}}, \bibinfo {author} {\bibfnamefont {Y.}~\bibnamefont {Shen}}, \bibinfo {author} {\bibfnamefont {K.}~\bibnamefont {Luo}}, \bibinfo {author} {\bibfnamefont {Q.}~\bibnamefont {An}}, \bibinfo {author} {\bibfnamefont {P.}~\bibnamefont {Gao}}, \bibinfo {author} {\bibfnamefont {P.}~\bibnamefont {Xiao}},\ and\ \bibinfo {author} {\bibfnamefont {Y.}~\bibnamefont {Zou}},\ }\bibfield  {title} {\bibinfo {title} {Harnessing dislocation motion using an electric field},\ }\href {https://doi.org/10.1038/s41563-023-01572-7} {\bibfield  {journal} {\bibinfo  {journal} {Nature Materials}\ }\textbf {\bibinfo {volume} {22}},\ \bibinfo {pages} {958–963} (\bibinfo {year} {2023})}\BibitemShut {NoStop}%
\bibitem [{\citenamefont {Alfonzetti}\ \emph {et~al.}(2006)\citenamefont {Alfonzetti}, \citenamefont {Dilettos},\ and\ \citenamefont {Salerno}}]{SimulatedAnnealingResetAlfonzetti2006}%
  \BibitemOpen
  \bibfield  {author} {\bibinfo {author} {\bibfnamefont {S.}~\bibnamefont {Alfonzetti}}, \bibinfo {author} {\bibfnamefont {E.}~\bibnamefont {Dilettos}},\ and\ \bibinfo {author} {\bibfnamefont {N.}~\bibnamefont {Salerno}},\ }\bibfield  {title} {\bibinfo {title} {Simulated annealing with restarts for the optimization of electromagnetic devices},\ }\href {https://doi.org/10.1109/tmag.2006.871675} {\bibfield  {journal} {\bibinfo  {journal} {IEEE Transactions on Magnetics}\ }\textbf {\bibinfo {volume} {42}},\ \bibinfo {pages} {1115–1118} (\bibinfo {year} {2006})}\BibitemShut {NoStop}%
\bibitem [{\citenamefont {Debenedetti}\ and\ \citenamefont {Stillinger}(2001)}]{GlassTransitionDebenedettiNature2001}%
  \BibitemOpen
  \bibfield  {author} {\bibinfo {author} {\bibfnamefont {P.~G.}\ \bibnamefont {Debenedetti}}\ and\ \bibinfo {author} {\bibfnamefont {F.~H.}\ \bibnamefont {Stillinger}},\ }\bibfield  {title} {\bibinfo {title} {Supercooled liquids and the glass transition},\ }\href {https://doi.org/10.1038/35065704} {\bibfield  {journal} {\bibinfo  {journal} {Nature}\ }\textbf {\bibinfo {volume} {410}},\ \bibinfo {pages} {259–267} (\bibinfo {year} {2001})}\BibitemShut {NoStop}%
\bibitem [{\citenamefont {Nesbitt}\ and\ \citenamefont {Hebard}(2007)}]{GlassyJunctionsPhysRevB}%
  \BibitemOpen
  \bibfield  {author} {\bibinfo {author} {\bibfnamefont {J.~R.}\ \bibnamefont {Nesbitt}}\ and\ \bibinfo {author} {\bibfnamefont {A.~F.}\ \bibnamefont {Hebard}},\ }\bibfield  {title} {\bibinfo {title} {Time-dependent glassy behavior of interface states in $\mathrm{Al}\text{\ensuremath{-}}\mathrm{Al}{\mathrm{o}}_{x}\text{\ensuremath{-}}\mathrm{Al}$ tunnel junctions},\ }\href {https://doi.org/10.1103/PhysRevB.75.195441} {\bibfield  {journal} {\bibinfo  {journal} {Phys. Rev. B}\ }\textbf {\bibinfo {volume} {75}},\ \bibinfo {pages} {195441} (\bibinfo {year} {2007})}\BibitemShut {NoStop}%
\end{thebibliography}%

\section*{Acknowledgments}
The structural and chemical characterization and analysis is based upon work supported by the U.S. Department of Energy, Office of Science, National Quantum Information Science Research Centers, Superconducting Quantum Materials and Systems Center (SQMS) under contract number DE-AC02-07CH11359.
We acknowledge the support of Anna Grassellino and Akshay Murthy. All resistance measurements were conducted on FormFactor automated probe stations with the help of Anwar Aslam, Connor Smith, and Brandon Boiko.  We thank Hilal Cansizoglu and the Rigetti fab team members for junction process development and samples, as well as Rory Cochrane and Cassidy Berk for assistance with cryogenic measurements and TLS analysis. All data is available upon request to the corresponding author. 

\change
\section*{Author information}
\noindent These authors contributed equally: David~P.~Pappas, Mark~Field, Cameron~J.~Kopas, Joel~Howard, Xiqiao~Wang, and Lin~Zhou.
\\
\\
\noindent These authors jointly supervised this work: Andrew~Bestwick, Kameshwar~Yadavalli, Matthew~J.~Kramer, and Josh~Y. Mutus.\\
\\
\noindent Authors and Affiliations\\
\noindent\textbf{Rigetti Computing, 775 Heinz Avenue, Berkeley, CA 94710, USA}\\
\noindent David~P.~Pappas, Mark~Field, Cameron~J.~Kopas, Joel~A.~Howard, Xiqiao~Wang, Ella~Lachman, Kameshwar~Yadavalli, Alysson~Gold, Gregory~M.~Stiehl, Eyob~A.~Sete, Andrew~Bestwick ~and Josh~Y.~Mutus\\
\\
\noindent\textbf{Ames National Laboratory, Ames, IA 50011, USA}\\
\noindent Lin~Zhou, Jinsu~Oh, Matt~J.~Kramer\\
\\
\noindent\textbf{Department of Materials Science and Engineering, Iowa State University, Ames, IA 50011, USA}\\
\noindent Lin~Zhou, Jinsu~Oh, Matthew~J.~Kramer\\
\\
Contributions\\
D.P.P. designed the experiment and wrote the manuscript; M.F and X.W. made ABAA measurements; C.J.K. identified and provided samples and qubit measurement results; J.A.H. conducted analysis and frequency predictions; E.S. provided theoretical support; E.L. and A.G. conducted literature searches and provided critical input for the manuscript; L.Z. and J.O. conducted and interpreted microscopy; G.M.S. conducted and interpreted qubit measurements; A.B., K.Y and M.J.K. supervised the work; J.Y.M. motivated and supervised the work. All authors discussed the results.\\
\\
Corresponding Authors\\
David~P.~ Pappas~dpappas@rigetti.com or Josh~Y.~Mutus~jmutus@rigetti.com\\
\\
\section*{Ethics declarations}
\noindent Competing interests\\
D. P. Pappas, M. Field, C. Kopas, J. A. Howard, X. Wang, E. Lachman, K. Yadavalli, Allyson Gold, E. A. Sete, A. Bestwick and J.Y. Mutus are, have been, or may in the future be participants in incentive stock plans at Rigetti Computing Inc. D. P. Pappas, M. Field, X. Wang, E. A. Sete, and J.Y. Mutus are inventors on a pending patent applications related to this work (No. 63/612,176). The other authors declare that they have no competing interests.

\stopchange

\end{document}